\documentclass[letterpaper,twocolumn,10pt]{article}
\usepackage[dvipsnames]{xcolor}
\usepackage{usenix,epsfig,endnotes}
\usepackage{array,color,url}
\usepackage[linesnumbered,ruled,vlined]{algorithm2e}
\usepackage{indentfirst}
\usepackage{algpseudocode}
\usepackage{tabu}
\usepackage{amsmath}
\usepackage{amsthm}
\usepackage{amssymb}
\usepackage{bm}
\usepackage{bbm}
\usepackage{graphicx}
\usepackage{float}
\usepackage{epsfig}
\usepackage{subcaption}
\usepackage{epstopdf}
\usepackage{color}
\usepackage{comment}
\usepackage{booktabs}
\usepackage{multirow}
\usepackage{url}
\usepackage{caption}
\usepackage{verbatimbox}
\usepackage{comment}
\usepackage{diagbox}
\usepackage{enumitem}
\usepackage{setspace}
\usepackage[explicit]{titlesec}
\setlist[itemize]{leftmargin=*}
\usepackage{appendix}
\usepackage{color}
\usepackage{makecell}
\usepackage{bbm}
\usepackage{mathtools}
\usepackage{tikz}

\SetCommentSty{mycommfont}

\newtheorem{theorem}{Theorem}

\newtheorem{definition}{Definition}
\title{Data Poisoning Attacks to Locally Differentially Private Range Query Protocols}

\author{
{\rm Ting-Wei Liao$^{\dagger *}$}
\and
{\rm Chih-Hsun Lin$^\dagger$}
\and
{\rm Yu-Lin Tsai$^\dagger$}
\and
{\rm Takao Murakami$^{*\Psi}$}
\and
{\rm Chia-Mu Yu$^\dagger$}
\and
{\rm Jun Sakuma$^{*\partial}$}
\and
{\rm Chun-Ying Huang$^\dagger$}
\and
{\rm Hiroaki Kikuchi$^\wp$} \\
\and
$^\dagger$National Yang Ming Chiao Tung University
\and
$^*$RIKEN AIP
\and
$^\Psi$ISM and AIST
\and
$^\partial$Institute of Science Tokyo
\and
$^\wp$Meiji University\\
}

\date{}
\begin{document}
\maketitle

\newcommand{\citehere}{\textcolor{red}{[citehere]}}
\newcommand{\refhere}{\textcolor{red}{[refhere]}}
\newcommand{\continuehere}{\textcolor{red}{To Be Continue}}
\newcommand{\todo}{\textcolor{red}{TODO}}

\thispagestyle{empty}

\subsection*{Abstract}
Local Differential Privacy (LDP) has been widely adopted to protect user privacy in decentralized data collection. However, recent studies have revealed that LDP protocols are vulnerable to data poisoning attacks, where malicious users manipulate their reported data to distort aggregated results. In this work, we present the first study on data poisoning attacks targeting LDP range query protocols, focusing on both tree-based and grid-based approaches. We identify three key challenges in executing such attacks, including crafting consistent and effective fake data, maintaining data consistency across levels or grids, and preventing server detection. To address the first two challenges, we propose novel attack methods that are provably optimal, including a tree-based attack and a grid-based attack, designed to manipulate range query results with high effectiveness. \textbf{Our key finding is that the common post-processing procedure, Norm-Sub, in LDP range query protocols can help the attacker massively amplify their attack effectiveness.} In addition, we study a potential countermeasure, but also propose an adaptive attack capable of evading this defense to address the third challenge. We evaluate our methods through theoretical analysis and extensive experiments on synthetic and real-world datasets. Our results show that the proposed attacks can significantly amplify estimations for arbitrary range queries by manipulating a small fraction of users, providing 5-10x more influence than a normal user to the estimation. 

\section{Introduction}\label{sec: Introduction}

Local Differential Privacy (LDP)~\cite{duchi2013local,duchi2014privacy,kasiviswanathan2011can}, a decentralized variant of Differential Privacy (DP)~\cite{Dwork2014TheAF}, safeguards end-user privacy without relying on a trusted server. LDP guarantees that even data aggregators cannot deduce the exact values of users' private information. This approach has been widely adopted by leading companies, including Microsoft~\cite{ding2017collecting}, Apple~\cite{AppleDP}, and Google~\cite{Erlingsson2014RAPPORRA}. 

Among the various LDP applications, range query protocols play a critical role in enabling numerous real-world tasks ~\cite{downstream1_GPS,downstream_energy_auction,downstream_health_care}. For instance, mobile applications may perform range queries on geographical information (e.g. latitude, longitude) to collect users' geographical distribution, while mobile users want to prevent giving away their exact location.

Recently, data poisoning attacks have emerged as a significant threat to LDP protocols, with studies revealing their vulnerability to manipulated user data. For example, Cao et al.~\cite{Cao2021poisonFE} introduced the first targeted data poisoning attacks for LDP frequency estimation protocols, inflating target item frequencies to distort both frequency estimation and heavy-hitter detection. Subsequently, Wu et al.~\cite{Wu2022poisonKey} and Tong et al.\cite{itemset} demonstrated attacks on LDP key-value and LDP frequent itemset mining protocols, respectively. Li et al.~\cite{Li2023Finegrained} proposed a fine-grained attack on LDP protocols for mean and variance estimation, enabling precise manipulation of statistical estimations. Despite these advancements, the security of more complex protocols, such as range query protocols, remains unexamined.

Attackers may attempt to exploit range query protocols for personal or organizational gain. For instance, an unreasonable increase in the number of people participating in an event conducted in a specific region may attract advertisements and sponsorship~\cite{locationadv}. Alternatively, an unreasonable increase in the number of purchases of an item with certain range-based attributes (pricing, production times e.t.c) can create an illusion of fake demand and affect marketing strategy~\cite{demandprice}.

In this work, we investigate data poisoning attacks on LDP range query protocols, including tree-based and grid-based methods. Unlike simpler protocols, range query protocols involve not only LDP estimation but also multiple layers of complex post-processing. We address three key technical challenges in designing such attacks:

\begin{itemize}[itemsep=2pt,topsep=0pt,parsep=0pt]
	\item To enhance the effectiveness of attacks on LDP range query protocols, the attacker needs carefully craft fake data through LDP frequency estimation protocols by manipulating fake users. This manipulation steers the post-processing of LDP range query protocols in the desired direction.

    \item The attacker must maintain data consistency. Inconsistent data across the 1-D and 2-D grids in grid-based approaches or across different levels in tree-based approaches can lead to attack failure.

    \item The attacker must consider potential defense from the server and prepare adaptive countermeasure. Potential detections and adaptive attacks must be considered to successfully launch an attack.
\end{itemize}

We observe that the key distinction between poisoning range queries and poisoning frequency estimation (or other LDP tasks) is that LDP range query estimation often relies on complex post-processing to improve accuracy. In particular, many protocols employ Norm-Sub or similar normalization techniques to ensure that frequencies align with the properties required by range queries. However, we discover that an attacker can leverage these normalization methods to strengthen the impact of poisoning by carefully crafting malicious inputs. Building on this insight, we propose novel data poisoning attacks against LDP range query protocols to address these challenges. Our key contributions include: 
\begin{itemize}[itemsep=2pt,topsep=0pt,parsep=0pt] 
\item Presenting the \emph{first} data poisoning attacks aimed at LDP range query estimation. 
\item Developing tree-based and grid-based attacks via carefully engineered data, along with theoretical proofs of optimality. 
\item Proposing a detection algorithm as a countermeasure, compelling attackers to switch to less effective strategies.
\item Introducing an adaptive attack strategy that circumvents the defense mechanism while preserving strong attack efficacy. 
\end{itemize}

We evaluate our proposed attacks through theoretical analysis and empirical experiments. Specifically, while previous work on related topics observed utility scaling linearly with the number of fake users added, our works reveals a nonlinear relationship, where a fake user can provide 5-10x more influence than a normal user to the estimation of query. In addition, we implement the attacks and test them on both synthetic and real-world datasets. The results show that our attacks outperform baseline methods in terms of effectiveness. Furthermore, we investigate potential defense strategies and their corresponding adaptive attacks, experimentally validating their effectiveness.

\section{Related Work}\label{sec:related work}
\textbf{LDP on Frequency Estimation.}
Wang et al.~\cite{Wang2017LocallyDPFE} propose several methods for frequency estimation under Local Differential Privacy (LDP)~\cite{duchi2013local,duchi2014privacy,kasiviswanathan2011can}. Among these, Optimal Unary Encoding (OUE) is a widely used protocol. It encodes user data into a one-hot vector and perturbs the bits with specific probabilities to ensure privacy. In contrast, Optimal Local Hashing (OLH) uses a hash function and a key to encode user data, followed by random perturbation to achieve privacy guarantees. Typically, OUE is preferred for a small number of classes, while OLH is more efficient for larger class sizes due to its better communication efficiency.

\textbf{LDP on Range Query Estimation.} 
Numerous works on LDP range query estimation have been proposed and their methods vary greatly. Our paper mainly focus on AHEAD~\cite{AHEAD} and HDG~\cite{yang2020answeringmultidimensionalrangequeries}, representative for tree-based and grid-based approaches, respectively. AHEAD, a tree-based approach, uses interval trees to estimate range queries. AHEAD employs adaptive decomposition and a dynamic tree structure. By estimating frequencies level by level using OUE, the adaptive decomposition helps reduce errors caused by LDP noise. In contrast, HDG, a grid-based approach, uses 1- and 2-dimensional grids. This protocol collects data through OLH and stores it in grids. Range queries are answered by generating response matrices that leverage the information stored in these grids. In general, the tree-based approach is preferred in 1-D query, while the grid-based approach is preferred in multidimensional query~\cite{privnud}.

Due to the high variety of LDP range query protocols, we select AHEAD~\cite{AHEAD} and HDG~\cite{yang2020answeringmultidimensionalrangequeries} as our targets. We choose AHEAD as the representative for tree-based approaches due to its simplicity and generality. Many designs of AHEAD such as adaptive decomposition and user partitioning are later adopted by subsequent protocols. On the other hand, HDG is chosen as the representative for grid-based approaches because their response matrix generation algorithm is widely accepted as one of the best ways to estimate multi-dimensional range query~\cite{yang2020answeringmultidimensionalrangequeries,privnud,pripl}. Although our attack focuses on AHEAD ~\cite{AHEAD} and HDG ~\cite{yang2020answeringmultidimensionalrangequeries}, we tested our attack algorithms on other LDP range query protocols, including DHT~\cite{dht}, PrivNUD~\cite{privnud}, PriPL~\cite{pripl}. We particularly note that PRISM~\cite{prism} is not included in our comparison because we provide a formal proof in Appendix~\ref{appendix:prism} showing that PRISM does not satisfy $\epsilon$-LDP.

\textbf{Data Poisoning Attack.} 
LDP aggregators are susceptible to data poisoning attacks. For example, Cao et al.~\cite{Cao2021poisonFE} and Cheu et al.~\cite{cheu2021manipulation} are the first to study the poisoning attacks for LDP protocols. In particular, Cao et al.~\cite{Cao2021poisonFE} propose MGA as a targeted attack for LDP frequency estimation. MGA can increase the frequency of attacker-selected target items and promote them to be heavy hitters.

For LDP on key-value, Wu et al.~\cite{Wu2022poisonKey} developed targeted attacks to increase the frequency and mean of target keys. For LDP on mean and variance estimation, Li et al.~\cite{Li2023Finegrained} proposed fine-grained attacks to manipulate target items' mean and variance. In addition, Tong et al.~\cite{itemset} proposed data poisoning attack to frequent itemset mining where their attack manipulate top-k frequent itemset mining. As of now, there are no existing study about data poisoning on range query protocols.

\textbf{Defense to Data Poisoning Attack.} Various defenses have been proposed to counter poisoning attacks in LDP frequency estimation. For instance, LDPRecover~\cite{ldprecover} aims to reconstruct genuine frequencies by removing those contributed by fake users, while LDPGuard~\cite{ldpguard} effectively mitigates three particular data poisoning attacks. Song et al.~\cite{frequencydefenseefficient} propose a defense against output poisoning attacks (OPA) by employing additional cryptographic mechanisms. Moreover, Murakami et al.~\cite{suffledefense} investigate defenses within the shuffle model~\cite{Prochlo,SODA}. 

\section{Background}\label{sec:background} 

\subsection{Local Differential Privacy (LDP)}
In LDP, a randomized mechanism $M$ is employed to provide privacy for users' local data. A central server collects these perturbed data to reconstruct aggregated statistics. 

\begin{definition}[Local Differential Privacy]
A randomized mechanism $M: \mathcal{D} \rightarrow \mathcal{R}$ satisfies $\epsilon$-Local Differential Privacy (i.e., $\epsilon$-LDP) if and only if, for any pair of inputs $u, v \in \mathcal{D}$ and any subset of all possible outputs $R \subseteq \mathcal{R}$, we have  
\begin{align}
\nonumber
     Pr\left[M\left(u\right)\in R\right]\leq e^\epsilon Pr\left[M\left(v\right)\in R\right].
\end{align}
\end{definition}

\subsection{Frequency Estimation in LDP}
In the context of LDP range query protocols, frequency oracle (FO) are crucial for estimating the distribution of users' data across various dimensions. Plenty of methods \cite{AHEAD,yang2020answeringmultidimensionalrangequeries,privnud,pripl} have employed some FO to gain frequency information. Depending on the range query method, two popular FOs are commonly employed.

\textbf{Optimal Unary Encoding (OUE).} Given $d$ categories, OUE represents a category $v$ as a $d$-bits vector, where only the bit corresponding to $v$ is set to 1. OUE then perturbs all bits in the vector as follows:
\begin{align*}
    Pr[b'_i = 1] = \begin{cases}
        p, & \text{ if } b_i = 1,\\
        q, & \text{ if } b_i = 0,
    \end{cases}
\end{align*}
where $p = \frac{1}{2}$ and $q = \frac{1}{e^\epsilon + 1}$. After collecting all the bit vectors, $B$, from users, the server calculates the unbiased estimate of the frequency of a category $v$ as 
\begin{equation}\label{eq:OUE}
    \hat{f}_v = \frac{\sum_{i=1}^N B_i[v] - Nq}{N(p - q)}.
\end{equation}

\textbf{Optimal Local Hash (OLH).} OLH is typically used to handle a large number of categories. OLH constructs a universal hash family $\mathcal{H}$ that maps the original domain $[|\mathcal{D}|]$ to a smaller domain $[g] = \mathcal{C}$, where $g = e^\epsilon + 1$. Users independently and randomly select a hash function $h \in \mathcal{H}$ and report the hash function along with a perturbed value, $(h, y)$, based on the following:
\begin{equation}
    Pr[y = i] = \begin{cases}
        \frac{1}{2}, & \text{ if } h(v) = i, \\
        q, & \text{ if } h(v) \neq i,
    \end{cases}
\end{equation}
where $q = \frac{1}{e^\epsilon + 1}$. In our paper, we call the pair of a hash function and a value, $p = (h,y)$, to be a \textbf{hash pair}. We call the \textbf{support} of a hash pair, $S_p$, to be the set of items that $h$ hashes to the value $y$.

The server then estimates the frequency of category $v$ as:
\begin{equation}
    \hat{f}_v = \frac{\sum_{i=1}^N \mathbbm{1}_{v \in S_{p_i}} - Nq}{N(\frac{1}{2} - q)},
\end{equation}
where $\mathbbm{1}_{v \in S_{p_i}} = 1$ if and only if $h_i(v) = y_i$ where $p_i = (h_i, y_i)$.

The main difference between OUE and OLH lies in their communication costs. For large $d$, OUE incurs a larger communication cost of $\Theta(d)$, as it requires reporting $d$ bits of information. In contrast, OLH has a communication cost of $\Theta(\log N)$, because the size of the hash family is $\Theta(N)$, and returning the index of a hash function costs $\Theta(\log N)$. Despite this difference, their performance in terms of estimation variance is similar.

\subsection{Range Query}
In the range query problem, we consider a set of attributes $\mathcal{A}$, where each attribute $A_i \in \mathcal{A}$ represents a dimension and has domain $[c] = \{0, 1, \dots, c-1\}$. The length of the domains is set to be the same across all attributes, as padding dummy values into shorter domains is always possible. Each data point (or user in the LDP scenario) will hold a $|\mathcal{A}|$-dimensional vector $v \in [c]^{|\mathcal{A}|}$.

A $d$-dimensional range query $q$ concerns a size-$d$ subset of attributes $A_q \subseteq \mathcal{A}$, which is the set of dimension that the query is asking about. For each attribute $A_i \in A_q$, there is a range $[l_i, r_i)$ such that $0 \leq l_i < r_i \leq c$. The frequency of the range query $q$ is the proportion of data points that fall inside the range. Specifically, let $X$ be the set of data points, and let $X_i^j \in [c]$ be the value of the $i$-th data point for attribute $A_j$. The frequency of $q$, denoted $f_q$, can be calculated as:
\begin{equation}
    f_q = \frac{\sum_{X_i \in X}\prod_{A_j \in A_q} \mathbbm{1}_{X_i^j \in [l_j, r_j)}}{|X|},
\end{equation}
where $\mathbbm{1}_{X_i^j \in [l_j, r_j)} = 1$ if and only if $X_i^j \in [l_j, r_j)$.

\subsection{Range Query Estimation in LDP}
In general, LDP range query protocols consist of three steps: frequency estimation, post-processing, and query estimation. These protocols often strike a balance between LDP estimation error from frequency estimation and uniformity error, which arises from assuming a uniform underlying user distribution. Different protocols may adopt varying methods for these steps depending on their data structures and error analysis. In this paper, we focus on two victim protocols, AHEAD~\cite{AHEAD} and HDG~\cite{yang2020answeringmultidimensionalrangequeries}, which are a tree-based and a grid-based approach, respectively. We run our proposed attack on the other range query protocols~\cite{dht, privnud, pripl} to demonstrate that the attack effect can be transferred (see Section~\ref{exp:other_protocol}). 

\subsubsection{Tree-based Approach~\textemdash~AHEAD~\cite{AHEAD}} 
To begin with, AHEAD adaptively decomposes the domain range into smaller ranges iteratively. In each step, AHEAD estimates the frequencies of the lowest layer (leaf nodes) using OUE. After the estimation, AHEAD examines each leaf node and decides whether to split it into smaller ranges based on its frequency.

Once the tree structure is complete, AHEAD performs post-processing to remove negative frequencies and ensure parent-child consistency. To address negative frequencies, AHEAD applies a normalization method called \textbf{Norm-Sub}~\cite{Normsub} layer by layer. In each layer, Norm-Sub transforms the vector of node frequencies into a non-negative, sum-to-one frequency vector. Specifically, for a frequency vector $F = \langle f_1, f_2, \dots, f_k \rangle$, Norm-Sub finds a real-valued threshold $\delta$ such that
\begin{equation}\label{eq:normsub}
    \sum_{i=1}^k \max(f_i - \delta, 0) = 1,
\end{equation}
and returns $F' = \text{Norm-Sub}(F) = \langle f'_1, f'_2, \dots, f'_k \rangle$, where $f'_i = \max(f_i - \delta, 0)$ for all $1 \leq i \leq k$.

In addition, to maintain the consistency between the parent and child nodes' frequencies, AHEAD performs the following operation in a bottom-up fashion:
\begin{small}\begin{equation}\label{eq: tree_consistency}
    \Tilde{f}(n) = \begin{cases}
        \lambda \hat{f}(n) + (1 - \lambda)\sum_{v \in \text{Child}(n)} \Tilde{f}(v), & \text{ if } n \text{ is internal node}, \\
        \hat{f}(n), & \text{ otherwise},
    \end{cases}
\end{equation}
\end{small}
where $\Tilde{f}$ and $\hat{f}$ represent the frequency after and before consistency, respectively. The parameter $\lambda = \frac{\text{Var}(\text{Child}(n))}{\text{Var}(\text{Child}(n)) + \text{Var}(n)}$ is inversely proportional to the variance of the estimation. In practice, $\lambda$ is set to $\frac{\# \text{ of children}}{\# \text{ of children} + 1}$, assuming a uniform distribution in user data.

The post-processing is performed once in that order, and finally, in query estimation, AHEAD sums the frequencies of the nodes corresponding to the query. See \ref{fig:AHEAD} for the explanation of general algorithm flow of AHEAD.

\subsubsection{Grid-based Approach~\textemdash~HDG ~\cite{yang2020answeringmultidimensionalrangequeries}}
In HDG, ${d \choose 2}$ 2-dimensional grids and $d$ 1-dimensional grids are prepared for the total of $d$ dimensions. Each 1-D grid and 2-D grid splits the original domains into $g_1$ cells and $g_2$ columns, respectively. HDG then divides all users into $d + {d \choose 2}$ groups and estimates each grid separately using OLH.

To ensure consistency in HDG's post-processing, HDG uses the redundancy among the grids to reduce estimation variance. In particular, there are several 1-D grid cells and multiple 2-D grid columns corresponding to the same domain range. Consistency check is done by calculating the weighted average of those columns and cells and updating them. Specifically, for every grid $G_i$, let $S_i = \frac{g_1}{g_2}$ if $G_i$ is 1-D and $S_i = g_2$ if it is 2-D. HDG divides the original domain of each dimension into $g_2$ fractions. For every fraction $c_j$, let $F_{G_i, c_j}$ represent the sum of the frequencies corresponding to $c_j$ in $G_i$. Note that there are one 1-D grid and $d - 1$ 2-D grids that contribute to the frequency for $c_j$. Consistency on $c_j$ is ensured by adding the frequency of each cell in $G_i$ with $(F_{c_j} - F_{G_i, c_j})/S_i$, where $F_{c_j} = \left(\sum_{i} \frac{F_{G_i, c_j}}{S_i}\right)/\left(\sum_{i} 1/S_i\right)$.

To remove negativity, HDG adopts the same method as AHEAD. It applies the Norm-Sub operation on each grid. Finally, in query estimation, HDG references the 1-D and 2-D grid frequencies for every pair of dimensions in the query to construct a response matrix. Treating them as answers for 2-D queries, HDG then constructs the answer for a multidimensional query using response matrices. The detailed algorithm can be found in ~\cite{yang2020answeringmultidimensionalrangequeries}.

\section{Threat Model}\label{sec: threat model}
We delineate our threat model with respect to attacker's knowledge, capability, and goal.

\textbf{Attack Knowledge and Capability.}
We consider a scenario in which $N$ genuine users and an untrusted server jointly run an LDP range query protocol (e.g., AHEAD or HDG). Our attacks hinge on the following assumptions, consistent with prior works~\cite{Cao2021poisonFE, Wu2022poisonKey, itemset, Li2023Finegrained}. 

First, the attacker can inject $M<N$ fake users who submit carefully fabricated data to the server. The attacker exerts full control over these fake users (e.g., by creating fake accounts) and can instruct them to generate arbitrary data. Accordingly, the server computes range query results using reports from all $N+M$ users . Second, the attacker knows which LDP protocol is in use, enabling the crafted fake data to conform precisely to the expected format.

Together, these two assumptions facilitate an \emph{output poisoning attack} (OPA), a setting commonly examined in the literature~\cite{Cao2021poisonFE, Wu2022poisonKey, itemset, Li2023Finegrained} and the main focus of this paper. Third, the attacker can estimate $N$, as the number of real users is often publicly disclosed or inferable~\cite{Erlingsson2014RAPPORRA,Li2023Finegrained,windows,microsoftreveal}, albeit imperfectly. As a result, we investigate in Section~\ref{exp:treemismatchN} how inaccuracies in this estimate affect attack performance.

It is evident that fake accounts are both prevalent and inexpensive to obtain~\cite{bay2019social,fakeuserprice,thomas2013trafficking}, allowing $M$ to be quite large. For example, fake Facebook or X (Twitter) accounts cost around \$0.3 each~\cite{fakeuserprice}, while phone-verified Google accounts can be acquired for about \$0.03~\cite{thomas2013trafficking}, making it straightforward to amass a sufficient number of fake users. Nevertheless, $M$ cannot grow arbitrarily, as purchasing fake accounts still incurs costs. In practice, an attacker would acquire only as many as needed to strike a balance between the attack's performance and economic benefit. Throughout our experiments, we follow the standard practice of considering the ratio of fake users $\rho=M/(M+N)$ to lies within $\rho \in [0.05, 0.2]$\footnote{Although prior studies~\cite{Cao2021poisonFE, Wu2022poisonKey, itemset, Li2023Finegrained} have also considered \(\rho = \tfrac{M}{M+N} = 0.001\) or \(0.01\), such low proportions yield minimal attack effectiveness.}~\cite{Cao2021poisonFE, Wu2022poisonKey, itemset, Li2023Finegrained}. 

Moreover, since the attacker injects these fake users, all of them can be coordinated centrally, allowing the attacker to compute an overall poisoning strategy and direct each fake user’s outputs accordingly.

\textbf{Attack Goal.}
Our attack seeks to maximize the \textit{response} produced by the server when estimating the \textit{target range query $q_t$}, that is, to maximize the estimated LDP range query result for $q_t$. 

Let $\mathcal{M}_{X_{\text{real}} \cup X_{\text{fake}}}$ be a function that maps any multidimensional range query $q_t$ to a real value using an LDP range query protocol (e.g., AHEAD or HDG) to the data $X_{\text{real}}$ and $X_{\text{fake}}$, where $X_{\text{real}}$ ($X_{\text{fake}}$) denotes the report sent from real (fake) users. For simplicity, we refer to the reports sent by real (fake) users as \textit{real data} (\textit{fake data}). The attacker’s goal is to solve the following optimization problem:
\begin{equation}\label{eq: objective}
    \max_{X_{\text{fake}}} \mathcal{M}_{X_{\text{real}} \cup X_{\text{fake}}}(q_t).
\end{equation}

\textbf{Note.} In the design of our attack on HDG, we first optimize a simplified version of HDG protocol theoretically due to the high complexity of HDG for theoretical analysis. We simplify the protocol so that, when post-processing begins, HDG will run consistency and Norm-Sub only once. In other words, we remove the iterative post-processing for theoretical analysis. 

While removing iterative post-processing may increase the estimation variance of HDG, the overall working mechanism of the algorithm would remain unchanged. To also demonstrate the effectiveness of our designed attack, we will both verify our theoretical results on the simplified HDG and demonstrate the attack performance on the original HDG protocol in Section~\ref{sec: experiment}.

\section{Attacks} \label{sec: attacks}
Here, we focus our attack on AHEAD~\cite{AHEAD} and HDG~\cite{yang2020answeringmultidimensionalrangequeries}. Attacks for other methods could follow suit since AHEAD and HDG cover common ideas in LDP range query protocols. DHT~\cite{dht}, PrivNUD~\cite{privnud}, PriPL~\cite{pripl} will be tested in Section~\ref{exp:other_protocol} using the same attacks. 
\subsection{Strawman Methods}\label{sec:baseline}
We begin with a strawman attack by adopting MGA~\cite{Cao2021poisonFE} into attacking both AHEAD and HDG. The main idea lies in that range query estimation can naturally be considered an aggregation of multiple frequency estimations. By treating each point in the domain as an item in frequency estimation, we can apply MGA to boost the estimation of $q_t$ by designating all points within $q_t$ as the target items.

Specifically, this strawman attack applies MGA to OUE for attacking AHEAD, and MGA to OLH for attacking HDG. The details are presented in Algorithms~\ref{algo: MGA_on_Tree} and~\ref{algo: MGA_on_Grid} in Appendix~\ref{sec: Strawman Method: MGA}.

Although this strawman attack achieves some effectiveness in terms of poisoning (see Sections~\ref{sec: Impact of epsilon} and~\ref{sec: Impact of rho}), it can be perfectly detected by simple defenses (see Section \ref{exp:tree_detection}) because the number of 1s in OUE exhibit statistical abnormality. Particularly, MGA on tree will often target a large set of items for a moderate-sized query, making it easily detectable. On the other hand, it is as well unclear whether the strawman method could achieve optimality in terms of attack performance for the given goal of the attacker. 

As a result of vulnerabilities in both optimality and detection, we propose our attacks—attack on tree (AoT) and attack on grid (AoG)—which achieve provable optimality in terms of attack performance while ensuring stealthiness.


\subsection{Main Idea}
Our goal is to design an attack that maximizes $\mathcal{M}_{X_{\text{real}} \cup X_{\text{fake}}}(q_t)$ (Eq.~\ref{eq: objective}). However, the attacker can only inject fake data during the user's step of the protocol (i.e., the reports from either OUE or OLH). Subsequently, these fake data must pass through various post-processing steps before contributing to query estimation. Hence, the attacker is required to meticulously craft fake data such that it generates an effective response for the target query.

\begin{figure}[!hbt]
    \centering
    \begin{minipage}{0.23\textwidth}
    \includegraphics[width=\linewidth,height=73pt]{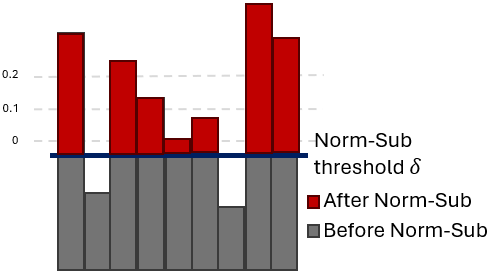}
    \caption{Norm-Sub}
    \label{fig:normsub}
    \end{minipage}\hspace{0.01\textwidth}\begin{minipage}{0.23\textwidth}
        \includegraphics[width=\linewidth,height=73pt]{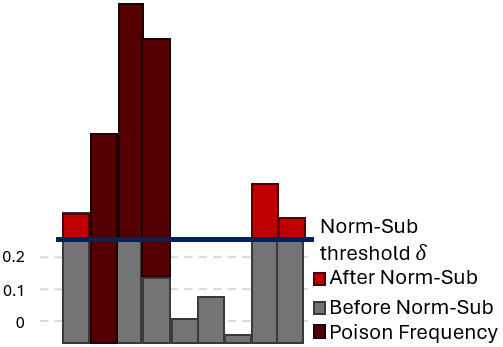}
    \caption{Poisoned Norm-Sub}
    \label{fig:normsubpoison}
    \end{minipage}
\end{figure}

The key insight behind our approach involves a common normalization technique, \emph{Norm-Sub}, widely adopted in many LDP range query protocols~\cite{AHEAD,yang2020answeringmultidimensionalrangequeries,privnud} for its simplicity and effectiveness. Norm-Sub normalizes frequencies by retaining only the top frequencies such that their total equals 1. Norm-Sub, as illustrated in Figure~\ref{fig:normsub} can be seen as plotting frequencies as a histogram and moving a horizontal bar from top to bottom until the area above the bar sums to 1, at which point the portion above the bar becomes the new frequencies. 

Given this post-processing mechanism, the attacker can inject a large number of higher frequencies which can effectively flatten out all other frequencies, preserving only those that serve the attacker’s purpose.

By exploiting Norm-Sub in such a manner, we can engineer the attack input to eliminate undesired frequencies to mount our attack successfully. See Figure~\ref{fig:normsubpoison} for the illustration of  how attackers are able to utilize it.

\begin{figure}[!hbt]
    \centering
    \begin{minipage}{0.18\textwidth}
    \includegraphics[width=\linewidth,height=73pt]{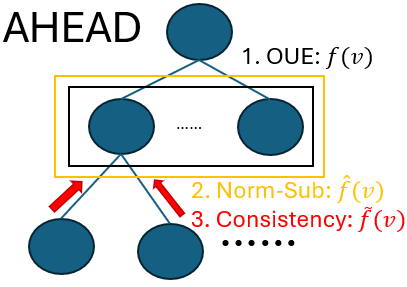}
    \caption{AHEAD}
    \label{fig:AHEAD}
    \end{minipage}\begin{minipage}{0.30\textwidth}
        \includegraphics[width=\linewidth,height=73pt]{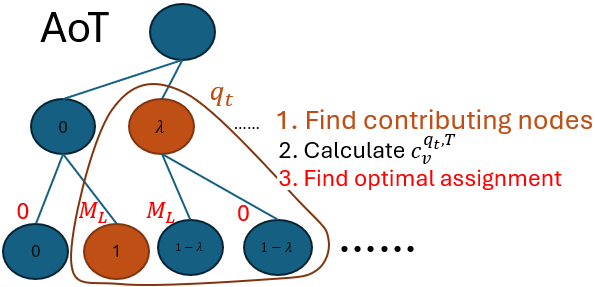}
    \caption{AoT procedure}
    \label{fig:AoTprocedure}
    \end{minipage}
\end{figure}
\subsection{Attacking AHEAD}\label{sec:aot}
We design our attack, Attack on Tree (AoT), on AHEAD to maximize the optimization goal of $\mathcal{M}_{X_{\text{real}} \cup X_{\text{fake}}}(q_t)$ (Eq.~\ref{eq: objective}). Let \(T\) be the tree structure constructed by AHEAD (implicitly dependent on both \(X_{\text{real}}\) and \(X_{\text{fake}}\)), and let \(V(T)\) be the set of nodes in \(T\). From Eq.~\ref{eq: objective} and AHEAD’s estimation, we derive 
\begin{equation}\label{eq: tree_anal_1}
    \mathcal{M}_{X_{\text{real}} \cup X_{\text{fake}}}(q_t) \;=\; \sum_{v \in V(T)} \Tilde{f}(v)\,\mathbbm{1}_{v \text{ is part of } q_t \text{ on } T},
\end{equation}
where \(\Tilde{f}(v)\) is the post-processed frequency of node \(v\) (using data from both \(X_{\text{real}}\) and \(X_{\text{fake}})\). Intuitively, this decomposes the query estimation into a sum over the nodes that constitute the queried interval. The indicator \(\mathbbm{1}_{v \text{ is part of } q_t \text{ on } T}\) is 1 if node \(v\) lies in the decomposition of \(q_t\) on \(T\), and 0 otherwise.

Considering AHEAD’s consistency steps (Eq.~\ref{eq: tree_consistency}), Eq.~\ref{eq: tree_anal_1} can be written as a linear combination of the \emph{pre-consistency} frequencies. Let \(\hat{f}(v)\) denote the frequency of node \(v\) before the consistency operation. Then,
\begin{equation}\label{eq: tree_anal_2}
    \sum_{v \in V(T)} \Tilde{f}(v)\,\mathbbm{1}_{v \text{ is part of } q_t \text{ on } T}
    \;=\;
    \sum_{v \in V(T)} c_v^{q_t,T}\,\hat{f}(v),
\end{equation}
where \(c_v^{q_t,T} \in \mathbb{R}^+ \cup \{0\}\) is a coefficient derived by substituting Eq.~\ref{eq: tree_consistency} into Eq.~\ref{eq: tree_anal_1}, i.e., reversing the consistency step of AHEAD. We name the coefficient \(c_v^{q_t,T}\) as \emph{tree coefficient}.

Since OUE and Norm-Sub operations are carried out in a layer-wised manner, Eq.~\ref{eq: tree_anal_2} can also be rearranged on a per-layer basis. Let \(L(T)\) be the set of layers in \(T\). We can derive
\begin{equation}\label{eq: obj_tree_layer}
    \sum_{v \in V(T)} c_v^{q_t,T} \,\hat{f}(v)
    \;=\;
    \sum_{L \in L(T)} \sum_{v \in L} c_v^{q_t,T} \,\hat{f}(v).
\end{equation}
Therefore, for each layer \(L\), the attacker can simply maximize 
\begin{equation}\label{eq:optimize_tree}
    \sum_{v \in L} c_v^{q_t,T} \,\hat{f}(v)
\end{equation}
to achieve the desired attack objective.

Maximizing Eq.~\ref{eq:optimize_tree} requires the attacker to carefully craft fake data that influences Norm-Sub and yields the desired values of \(\hat{f}(v)\). Suppose that the attacker controls \(M_L\) fake users at layer \(L\) where each user submits a bit for every node \(v \in L\) through OUE. The attacker thus is required to determine an \emph{assignment} \(A = \{a_v\}_{v \in L}\), where \(0 \le a_v \le M_L\) specifies how many fake users will report “present” for each node \(v\). 

To maximize attack performance in Eq. \ref{eq: objective}, we demonstrate in Theorem~\ref{thm:tree_optimal} that optimality can be achieved, with an \emph{optimal assignment} of a specific form constructed in Theorem~\ref{thm:tree_optimal}. Due to space constraint, we leave the proof of Theorem~\ref{thm:tree_optimal} in Appendix \ref{proof: tree_optimal}.

\begin{theorem}\label{thm:tree_optimal}
    Let \(L\) be a set of nodes, and let \(c_v\) be the tree coefficient for layer $L$. Arrange the nodes into \(v_1, v_2, \dots, v_{|L|}\) so that \(c_{v_1} \ge c_{v_2} \ge \dots \ge c_{v_{|L|}}\). Define an assignment \(A = \{a_i\}_{i=1}^{|L|}\) with \(0 \le a_i \le M_L\). There is an optimal assignment \(A^{\text{OPT}} = \{a_i^{\text{OPT}}\}_{i=1}^{|L|}\) and an integer \(k\) \((1 \le k \le |L|)\) that maximizes Eq.~\ref{eq:optimize_tree}. Moreover, \(A^{\text{OPT}}\) has the form
    \[
        a_i^{\text{OPT}}
        =
        \begin{dcases}
            M_L & \text{if } i < k,\\
            c   & \text{if } i = k,\\
            0   & \text{if } i > k,
        \end{dcases}
    \]
    where \(0 \le c < M_L\). We call the form \textbf{potential optimal} form.
\end{theorem}

\paragraph{Practical Considerations.} 
To apply Theorem~\ref{thm:tree_optimal} in practice, two main adjustments are needed. First, because attackers do not know the real users’ data \(X_{\text{real}}\), our algorithm assumes these data are uniformly distributed. This assumption is common in many LDP range query protocols~\cite{AHEAD,yang2020answeringmultidimensionalrangequeries,privnud}, including AHEAD and HDG. 

Second, while the optimal assignment can be computed for a fixed tree structure, in AHEAD and similar tree-based protocols, the tree itself can depend on the frequency inputs. Hence, our attack first simulates the protocol’s tree-building step under the assumption of uniform data, then computes an optimal assignment based on the resulting tree structure.

With these assumptions, the attacker can enumerate all potential optimal assignments, perform Norm-Sub, compute \(\sum_{v \in L} c_v^{q_t,T}\hat{f}(v)\), and select the potential optimal assignment that maximizes Eq.~\ref{eq:optimize_tree}. Theorem \ref{thm:tree_optimal} guarantees that this assignment will maximize Eq.~\ref{eq:optimize_tree} among all possible assignment. In practice, optimal assignment may contain one non-integer, and rounding the number creates negligible difference since the number of users is often large. In addition, to find an optimal assignment, a brute-force search takes \(O(M_L \,|L|^2)\) time, although careful pruning can improve this to \(\Theta(|L|^2 \log |L|)\) (see Appendix~\ref{sec:efficientsearch}).

In higher layers of the tree, some layer \(L\) may have zero tree coefficient \(c_v^{q_t,T}=0\) for all $v \in L$. Our proof does not dictate how to assign 1s or 0s for these nodes, as they do not affect the final frequency, but rather affect how the tree growth. To handle such cases, we evaluate three simple heuristics experimentally: (1) \emph{Zero}, assigning 0 to all nodes; (2) \emph{One}, assigning 1 to all nodes; and (3) \emph{Path}, assigning 1 only to those nodes along the path from target leaves to the root.

\paragraph{Algorithm.}  
Algorithm~\ref{algo: AoT} in Appendix~\ref{sec: Main Algorithms} summarizes our AoT against AHEAD. Lines \ref{algo: AoT_coef_start}--\ref{algo: AoT_coef_end} compute the coefficients \(c_v^{q_t,T}\) by traversing downward from the nodes used to estimate \(q_t\). Upon the server’s request for layer \(L\), the attacker incorporates \(L\) into the tree (line~\ref{algo: aot_getTree}), then sorts the nodes in \(L\) by \(c_v^{q_t,T}\) (line~\ref{algo: aot_getCoef}). If all \(c_v^{q_t,T}\) values are zero, a heuristic strategy is used. Otherwise, by Theorem~\ref{thm:tree_optimal}, an optimal assignment must lie within the enumerated set \(A\) (line~\ref{algo: AoT_all_ass}). The attacker exhaustively searches for the best assignment under Norm-Sub, selects \(a_{\text{best}}\) (lines~\ref{algo: aot_best}--\ref{algo: aot_return}), and finally makes the fake users submit their data according to \(a_{\text{best}}\). See Figure \ref{fig:AoTprocedure} for the flow of our algorithm.

\subsection{Attacking HDG}
Our objective is to inflate the estimated value of the target query under HDG (Eq.~\ref{eq: objective}). Recall that HDG estimates queries by constructing response matrices from grid frequencies. In the following, we show that these response matrices can be manipulated to yield the maximum possible estimate if the grid frequencies adopt a particular configuration.

\textbf{Conditions for Attack Optimality.}
To maximize the attack performance, we say that frequencies are \emph{fully concentrated} in \(q_t\) if, for every 1-D and 2-D grid, any cell lying outside the range \(q_t\) has frequency zero after post-processing. The theorem below demonstrates that an attack is optimal if it achieves such full concentration in \(q_t\). We leave the proof of Theorem~\ref{thm:fullycon} in Appendix~\ref{proof:fullycon}.

\begin{theorem}\label{thm:fullycon}
    In HDG, if the attacker chooses \(X_{\text{fake}}\) so that the frequencies are fully concentrated in \(q_t\) after post-processing, then the attack succeeds, i.e.
    \[
        \mathcal{M}_{X_{\text{real}} \cup X_{\text{fake}}}(q_t) \;=\; 1.
    \]
\end{theorem}

By Theorem~\ref{thm:fullycon}, the attacker’s goal reduces to constructing \(X_{\text{fake}}\) such that the grid frequencies become fully concentrated in \(q_t\). Unlike AHEAD, which relies on OUE, HDG leverages OLH. This implies that fake users cannot simply select any subset of cells to return to the server; instead, each fake user must choose a \emph{hash pair}\footnote{For LDP guarantees, the user should randomly choose their own hash function (rather than having it assigned by the server), maintaining the property of a universal hash family.} for every grid. As a result, identifying “good” hash pairs that achieve full concentration of frequencies in \(q_t\) would be crucial to the attacker.

\paragraph{On “Good” Hash Pairs for the Attacker.}
We define three constraints—\textit{Size}, \textit{Inclusive}, and \textit{Column constraint}—that characterize the “good” hash pairs the attacker needs to launch attack successfully. The first two constraints ensure that the hash pair has sufficiently large support while restricting all the support to lie entirely within $q_t$. The Column constraint ensures consistent support across multiple grids that share at least one dimension from \(q_t\). We state these formally in Theorem~\ref{thm: grid_constraint}, proved in Appendix~\ref{proof:grid_optimal}.

\begin{figure}
    \centering
    \includegraphics[width=\linewidth]{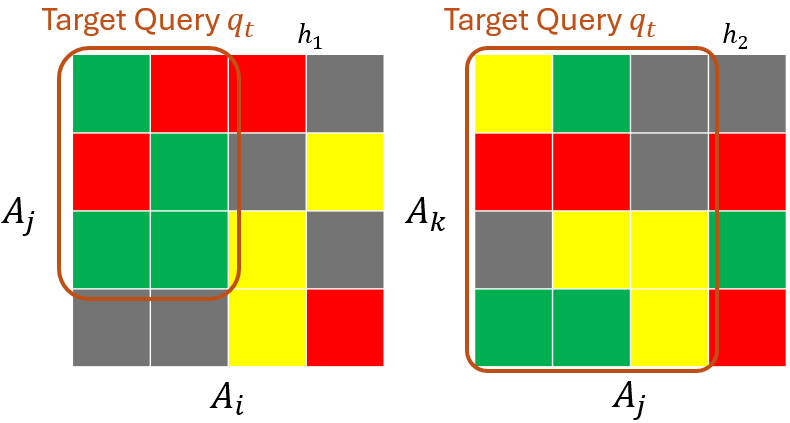}
    \vspace{-0.4cm}
    \caption{The figures depict two 2-D grids with attributes $(A_i, A_j)$ and $(A_j, A_k)$. Each color represents a distinct hash key, and every cell is assigned a color according to the hash function shown in the top-right corner. Given $w_2 \leq 4$, both the Inclusive and Size constraints are satisfied by using the hash pair $(h_1, \text{\textcolor{ForestGreen}{$\blacksquare$}})$ in the left grid and $(h_2, \text{\textcolor{Yellow}{$\blacksquare$}})$ in the right grid. Furthermore, the column constraint is met because each corresponding column range contains the same number of cells $(1,1,2,0)$ in the first, second, third, and fourth columns/rows, respectively.}
    \label{fig:grid_good_hash}
\end{figure}

\begin{theorem}\label{thm: grid_constraint}
    Let \(\rho = \frac{M}{N + M}\) be the fraction of malicious users. Let $g$ be the number of hash key in OLH. Denote by \(G_i^1\) a 1-D grid over dimension \(i\), and by \(G_{ij}^2\) a 2-D grid over dimensions \(i\) and \(j\). Denote by $g_1, g_2$ the granularity of 1-D and 2-D grids respectively. Let \(C(G)\) be the set of cells in grid \(G\), and let \(S_p(G)\) be the support of a hash pair \(p = (h, k)\), defined by \(S_p(G) = \{\,c \in C(G) \,\mid\, h(c) = k\}\).

    For each 1-D or 2-D grid \(G\), if the attacker can find a hash pair \(p_i\) or \(p_{ij}\) satisfying the constraints below, the attacker can fully concentrate frequencies in \(q_t\):

    \begin{enumerate}
        \item \textbf{Size constraint}: \(|S_{p}(G)| \ge w_1\) for 1-D grids or \(|S_{p}(G)| \ge w_2\) for 2-D grids, where
    \[
        w_1 
        = 
        \frac{\tfrac12 - \tfrac1g}{\rho} 
        \cdot
        \frac{(d-1)\,g_1 + g_2^2}{(d-1)\,(g_1 - 2g_2) + g_2^2},\]
        \[ w_2 
        = 
        \frac{\tfrac12 - \tfrac1g}{\rho} 
        \cdot
        \frac{g_2}{\,g_2 - 3 + \tfrac{3g_1}{g_1\,(d-1) + g_2^2}\,}.
    \]
        \item \textbf{Inclusive constraint}: Every cell \(c \in S_{p}(G)\) must lie within the range \(q_t\).
    \end{enumerate} 
    Furthermore, for any two 2-D grids \(G^2_{ij}\) and \(G^2_{ik}\) that share a dimension included in \(q_t\), the following condition must hold:

    \begin{enumerate}[start=3]
        \item \textbf{Column constraint}: Let these grids each have \(g_2\) columns corresponding to the same domain partitions \(D^i_1, D^i_2, \dots, D^i_{g_2}\). Denote by 
        \(\text{Col}_{ij}^r = S_{p_{ij}}(G^2_{ij}) \cap C\bigl(G^2_{ij}, D^i_r\bigr)\) 
        the set of cells in column \(r\) of \(G^2_{ij}\) that are in the support of \(p_{ij}\). We say \(p_{ij}\) and \(p_{ik}\) satisfy the Column constraint if, for every \(1 \le r \le g_2\),
        \begin{align}
            \bigl|\!\bigl|\text{Col}_{ij}^r\bigr| - \bigl|\text{Col}_{ik}^r\bigr|\bigr| \;\le\; 1.
        \end{align}

        A similar condition applies to pairs of grids \(G^1_i\) and \(G^2_{ij}\) sharing dimension \(i\). In that case, the same rule must hold for the rows or columns representing dimension \(i\) in both grids.
    \end{enumerate}
\end{theorem}
See Figure \ref{fig:grid_good_hash} for graphical example for three constraints.

\paragraph{Practical Considerations.}
The values \(w_1\) and \(w_2\) scale inversely with \(\rho\). Intuitively, the more fake users an attacker has, the looser the size constraints become, thus making it easier to find a suitable hash pair for each grid and achieve full frequency concentration in \(q_t\).

For a given grid \(G\), let \(c\) be the pessimistic estimation of probability that a randomly chosen hash pair \(p = (h,k)\) meets the Size, Inclusive, and Column constraints in \textit{the worst case}. Since \(\rho\) affects how much frequency the attacker can inject, an increase in \(\rho\) looses up Size constraint and thus raises \(c\). The probability of \emph{failing} to find any valid hash pair on \(G\) is then \((1 - c)^{|\mathcal{H}|\cdot g}\), where \(|\mathcal{H}|\) is the size of the hash family. Because \(|\mathcal{H}|\) is often very large (e.g., on the order of the number of users \cite{OLHhashnumber}), even modest improvements in \(c\) can push \((1 - c)^{|\mathcal{H}|\cdot g}\) close to 0, thus ensuring that a valid hash pair appears with high probability.

Given a tolerance \(\beta\) for failure (i.e., the probability that not all grids can be forced to fully concentrate in \(q_t\)), we want 
\begin{align}
    (1 - c)^{|\mathcal{H}|\cdot g}
    \;\le\;
    \frac{\beta}{\,\frac{d(d+1)}{2}\,},
\end{align}
so that across all \(d + \tfrac{d(d-1)}{2}\) grids (1-D and 2-D), the chance of finding a valid hash pair for each grid is at least \(1 - \beta\). 

In practice, one can estimate $c$ by enumerating the relevant column-constraint combinations. Under typical parameter settings (Section~\ref{sec: experiment}), the required $\rho$ based on the above calculation is approximately $0.301$. Nonetheless, empirically, AoG often succeeds in finding suitable hash pairs even when $\rho$ is below $0.2$ because our estimation of $c$ is pessimistic.

Consequently, given the extensive hash family available, it is reasonable to conclude that effective hash pairs are sufficiently abundant. With the calculated number of fake users, achieving full concentration in $q_t$ becomes feasible.

\paragraph{Algorithm.}
Algorithm~\ref{algo: AoG} in Appendix~\ref{sec: Main Algorithms} summarizes our proposed attack ob HDG, \emph{Attack on Grid} (AoG). Lines~\ref{algo: AoG_find_hash_start}--\ref{algo: AoG_find_hash_end} exhaustively search for a hash pair \((h,c)\) that satisfies the Inclusive, Size, and Column constraints in each grid. If no valid pair is found, the algorithm fails to guarantee full concentration in \(q_t\); as a fallback, it resorts to a heuristic approach (Algorithm~\ref{algo: AoG_heuristic} in Appendix~\ref{sec: Main Algorithms}) with slightly lower performance (Section~\ref{sec: experiment}).

Lines~\ref{algo: AoG_column_start}--\ref{algo: AoG_column_end} verify the Column constraint, ensuring that each newly chosen hash pair maintains consistency across different grids that share dimensions in \(q_t\). Instead of enumerating all compatible hash pairs across all grids simultaneously—which is computationally prohibitive—we adopt a greedy heuristic: for each grid in turn, we pick the first hash pair that satisfies the Inclusive, Size, and Column constraints, then update the column-constraint bookkeeping accordingly. This approach typically succeeds because (i) the set \(R_{q_t}(G)\) is often small, and (ii) valid hash pairs on the same grid typically share a similar shape.

Nonetheless, if no valid hash pair is found for some grids, or if the attacker lacks sufficient fake-user budget to meet \(w_1\) and \(w_2\), the \emph{heuristic AoG} (HAoG) algorithm offers a fallback. HAoG selects the hash pair that best approximates the size and inclusive constraints.

HAoG is presented in Algorithm~\ref{algo: AoG_heuristic} in Appendix~\ref{sec: Main Algorithms}. In Line~\ref{algo: AoG_heuristic_preference}, HAoG applies the function \(\textit{preference}(h,c,G)\) as a comparison key to select the maximum element in the set \(\mathcal{H} \times \mathcal{C}\). This preference function outputs two values: (i) the negative of the number of cells in \(S[h,c]\) that lie outside \(R_{q_t}\) (the \emph{primary} key), and (ii) the total number of cells in \(S[h,c]\) (the \emph{secondary} key). Intuitively, the first value captures how much \((h,c)\) violates the inclusive constraint (fewer out-of-range cells is better), while the second value indicates how well \((h,c)\) satisfies the size constraint (more in-range cells is better).

Moreover, in the 1-D preference function at Line~\ref{algo: AoG_heuristic_1D}, there is a factor of \(\frac{g_1}{g_2}\) in the denominator. Although this factor does not affect which element attains the maximum value, it becomes useful in the adaptive attack setting in Section~\ref{sec: Adaptive Attack}.

\section{Defense and Adaptive Attack}\label{sec: defense and adaptive attacks}

\subsection{Defense}\label{sec: defense}
To start with, we note that existing defenses are not effective in LDP range query protocols due to various reasons. For instance, LDPRecover\cite{ldprecover} has minimal effect on our attack and can significantly impair the estimation for both AHEAD and HDG when no attack is launched (See Appendix~\ref{sec: Detection Result via LDPRecover}). On the other hand, LDPGuard\cite{ldpguard} has an incompatible scenario with the LDP range query, designed specifically for three types of attacks on data poisoning for frequency estimation. Both \cite{suffledefense} and \cite{frequencydefenseefficient} require additional mechanisms beside original protocol. One requires a third-party shuffler and the other requires multiple rounds of cyptographic operations, both of which are incompatible to our attack scenerio. As a result, we propose our detection algorithm to detect attacks on LDP range query protocols.

A server aware of potential vulnerabilities may devise countermeasures to detect these attacks while preserving the protocol’s performance. An approach is to look for irregular data distributions by conducting hypothesis tests. We propose a defense strategy that targets both AoT and AoG, relying on hypothesis tests to identify statistical anomalies. In AoT, the key signal is a skew in the distribution of “1” bits returned through OUE, which may appear when an attack is underway. Meanwhile, in AoG, the maximum load (i.e., the number of times a hash function is chosen) can exhibit abnormal patterns. High maximum load may be observed when an attack on AoG occurs, while the real maximum load should remain relatively low if users honestly select hash function at random.

\textbf{Tree-based Defense.} In AHEAD, when the server requests an \(n\)-bit OUE, the number of 1s returned by a normal user follows the convolution of two distributions: \(\mathrm{Bin}(n-1, q)\) and \(\mathrm{Bin}\bigl(1,\tfrac12\bigr)\), where \(\mathrm{Bin}(m, p)\) denotes the binomial distribution with \(m\) trials and success probability \(p\).

We consider a detection mechanism, shown in Algorithm~\ref{algo: defense_tree} Appendix~\ref{sec: Detection Algorithm}, by identifying users who produce an excess or deficit of 1s. Specifically, the server defines the null hypothesis, \(H_0:\) \emph{each user follows the OUE protocol correctly}, and adopts a significance threshold \(\alpha\). Next, it defines a high-probability interval \(I = [a,b]\) (with \(0 \le a < b \le n\)) such that most legitimate users have a number of 1s in this interval. The test statistic is the count of users whose 1s lie outside of \(I\). If this count exceeds a threshold \(k\), we flag suspicious activity.

Intuitively, the server aims for a sensitive defense, i.e., one capable of detecting attacks even when \(M\) (the number of fake users) is relatively small. To achieve this, the defense converts \(\alpha\) to \(z_{\alpha}\), where \(\Pr[Z \le z_{\alpha}] = 1 - \alpha\) and \(Z\) stands for standard normal random variable. Let \(f_I\) be the probability that a normal user’s number of 1s lies within \(I\). We then define the threshold \(k\) as

\begin{equation}
    k \;=\; N_{\mathrm{OUE}} \,(1 - f_I)\;+\;z_{\alpha}\,\sqrt{\,N_{\mathrm{OUE}}\,(1 - f_I)\,f_I\,},
    \label{eq: tree_defense_minimize}
\end{equation} where $N_{\mathrm{OUE}}$ denotes the total number of users responding via OUE. By minimizing $k$ (equivalently, maximizing \(f_I\)), we can pin down the appropriate interval \(I\). 

\textbf{Grid-based Defense.} In OLH, each user is expected to randomly choose a hash function from \(\lvert\mathcal{H}\rvert\) possible functions. However, the attacker in AoG deliberately selects a small subset of hash functions and always reports them. This artificially skews the distribution of chosen hash functions, which can be detected through anomalies in the \emph{maximum load}, i.e., the count of how many times the most frequently chosen hash function appears.

Consider that \(\lvert\mathcal{H}\rvert\) hash functions exist and \(N_{\mathrm{OLH}}\) users respond with OLH. If every user truly picks a hash function uniformly at random, then modeling the situation as a balls-into-bins problem~\cite{ballsintobins,ballsbounds} suggests that the maximum load \(t\) behaves roughly like
\begin{align}
    t 
    \;=\; 
    \frac{\log N_{\mathrm{OLH}}}{\,\log N_{\mathrm{OLH}} - \log \lvert\mathcal{H}\rvert\,}.
\end{align}

In practice, \(t\) is typically small (e.g., under 40), even in the presence of millions of legitimate and fake users. Thus, the defense, as shown in Algorithm~\ref{algo: grid_based_defense} in Appendix~\ref{sec: Detection Algorithm}, can simply check whether the maximum load exceeds \(t\). The exact detection threshold can be fine-tuned via simulation for a given significance level~\(\alpha\).

\textbf{Summary of Defense.} Algorithms~\ref{algo: defense_tree} and~\ref{algo: grid_based_defense} essentially capture the abnormalities introduced by AoT and AoG. Empirically, these detection methods can identify AoT and AoG with over 40\% success. This observation motivates us to develop adaptive attacks that can completely evade detection. The experimental result can be found in Section \ref{exp:tree_detection}.

\subsection{Adaptive Attack}\label{sec: Adaptive Attack}
Attackers who anticipate potential defensive mechanisms may adapt their strategies to evade detection. Thus, we propose adaptive attacks for both tree-based and grid-based approaches, maximizing  attackers’ capabilities when countermeasures are active.

\textbf{Adaptive Tree-based Attack.}
For each fake user on a layer with $n$ nodes, assume that under AoT the user was originally instructed to report a vector $v$ containing $k$ ones. In our adaptive attack, Adaptive AoT (AAoT), shown in Algorithm~\ref{algo: adaptive_tree} in Appendix~\ref{sec: Adaptive Attack}, the user additionally samples a random variable $X \sim Bin(n-1, q) + Bin(1, p)$, which determines how many ones appear in the final submission. If $X > k$, the user randomly selects $X - k$ positions in $v$ that are $0$ and flips them to $1$. Otherwise, the user randomly selects $k - X$ positions in $v$ that are $1$ and flips them to $0$. By applying this step, no statistical difference in the number of $1$s is observed between fake and real data. As shown in Sections~\ref{sec: Impact of epsilon},~\ref{sec: Impact of rho}, and~\ref{exp:tree_detection}, the adaptive attack not only evades detection but also incurs minimal reduction in attack effectiveness.

\textbf{Adaptive Grid-based Attack.}
Suppose the server implements a defense with a detection threshold $t$, meaning it triggers whenever any hash function is selected more than $t$ times. The goal of our adaptive attack, Adaptive AoG (AAoG), illustrated in Algorithm~\ref{algo: adaptive_grid} in Appendix~\ref{sec: Adaptive Attack}, is for the attacker to choose a sufficiently large set of hash functions such that each one is used at most $t_L$ times by fake users, with $t_L \leq t$.

As in HAoG (Algorithm~\ref{algo: AoG_heuristic} in Appendix~\ref{sec: Main Algorithms}), the attacker prioritizes hash functions that best satisfy the size and inclusive constraints. However, in this adaptive scenario, each hash function is restricted to a single grid, and its compatibility may vary across different grids. Let $M_G$ be the number of fake users in grid $G$. Accordingly, the attacker must allocate a total of $\frac{M_G}{t_L}$ hash functions for grid $G$.

Although the approach remains largely heuristic, we could formulate this problem as a variant of the stable matching problem~\cite{Kleinberg+Tardos:06a}, matching hash functions to 1-D and 2-D grids. Each grid contains multiple possible hash functions, and the preference between a grid and a hash function is determined by that hash function’s compatibility with the grid’s size and inclusive constraints.

Furthermore, the attacker must decide on $t_L$ based on their risk tolerance. The detection mechanism can be triggered if genuine users happen to choose the same hash functions that fake users are selecting. Consequently, setting $t_L$ closer to $t$ raises the chance of activating the defense. To evaluate this risk, the attacker may run simulations to approximate the probability of detection for a chosen $t_L$ and threshold $t$.

Additionally, in Line \ref{algo: adaptive_grid_L}, the probability may also be computed through simulation. We will present our chosen value of $L$ in Section~\ref{sec: Setup}. The experimental results in Sections~\ref{sec:result} and~\ref{exp:tree_detection} demonstrate that the attack achieves high effectiveness while maintaining a low detection rate.

\def\syndataset{Gaussian }
\def\realonedataset{BlackFriday }
\def\realtwodataset{PaySim }
\section{Experiment} \label{sec: experiment}
\subsection{Setup}\label{sec: Setup}
\textbf{Dataset.} Our experiments use two synthetic datasets (Gaussian and Laplace) and five real-world datasets (BFive~\cite{bfive}, BlackFriday~\cite{BlackFriday}, Loan~\cite{loan}, PaySim~\cite{paysim}, and Salaries~\cite{salaries}). Appendix~\ref{sec: Dataset Description} provides details for each dataset.  
Due to space constraints, this section reports only the results for the BlackFriday dataset, which aligns with the incentive scenario in Section~\ref{sec: Introduction}. Results for the other datasets appear in the Appendix~\ref{sec:even_more_tree} and ~\ref{sec:even_more_grid}. For multi-dimensional queries, we filter out any ill-formed records before using the remaining data. For example, a total of 166 thousand records in BlackFriday remain for experimentation.

\textbf{Metric.} We measure attack performance with two metrics:
\begin{itemize}
    \item \textit{Response}: The estimated value of the target range query $q_t$ on an LDP range query protocol. 
    \item \textit{Efficiency}: Defined by $
        \mathcal{E} \;=\; \frac{\hat{f}_{q_t} \;-\; f_{q_t}}{\rho}$,
    where $f_{q_t}$ is the response under no attack for $q_t$ (real response), $\hat{f}_{q_t}$ is the poisoned response, and $\rho = \tfrac{M}{M+N}$.
\end{itemize}
These two metrics convey the same underlying information, but \textit{efficiency} highlights how effectively an attack leverages its fake users. In an ideal scenario with no variance, adding 1\% fake users would increase the response by 1\%, yielding $\mathcal{E} = 1$. However, due to the LDP mechanism and post-processing (particularly Norm-Sub), attackers can exploit protocol vulnerabilities to achieve higher $\mathcal{E}$. For instance, an attack with $\mathcal{E} = 5$ could raise the response from $0.1$ to $0.6$ while controlling only 10\% of the users. Besides, $\mathcal{E}$ can also evaluate other poisoning attacks. For example, in MGA (on OUE and OLH)~\cite{Cao2021poisonFE}, efficiency is bounded by $\tfrac{2 e^\epsilon}{\,e^\epsilon - 1\,}$ per target item.

\textbf{Parameter Setting.}
For AHEAD and HDG, we set the domain to $[c] = [0, 1023]$ and $[c] = [0, 63]$\footnote{For HDG, any query is automatically trimmed to fit the 2-D grid interval, ensuring consistency between 1-D and 2-D grids. An attack succeeding on the smaller, trimmed query also succeeds on the original, larger range.}, where the center of each query range is chosen uniformly from $[c]$, and the query length is chosen uniformly from $\bigl[\tfrac{c}{8}, \tfrac{3c}{8}\bigr]$. This setup produces moderately sized queries, preventing extreme cases. In AHEAD, we randomly pick 1 dimension out of 5 for 1-D queries; for HDG, we randomly pick 3 out of 5 dimensions for multi-dimensional queries. We generate 20 random queries and report the average response or efficiency.

Unless specified otherwise, we use $\epsilon = 1.0$ and $\rho = 10\%$ as defaults, with a server significance threshold $\alpha = 0.005$ for each OUE or OLH invocation. The number of hash functions equals the number of users in the same OLH round.\footnote{We assume the attacker cannot exploit this detail to learn the total number of users, since the server could use arbitrarily large values.} 

Finally, for AAoG, we estimate $L$ in Algorithm~\ref{algo: adaptive_grid} (Line~\ref{algo: adaptive_grid_L}) via simulation on different datasets. In BlackFriday, $L$ ranges from 8 to 10 depending on $\rho$.

\def\realonetreefolder{figures/trees/BlackFriday/RandomQuery/}
\def\realtwotreefolder{figures/trees/PaySim/RandomQuery/}

\subsection{Result}\label{sec:result}
\textbf{Tree.} Figures~\ref{fig:tree_differentrho_respond} and~\ref{fig:tree_differentrho_eff} illustrate our attacks’ effectiveness. Across these settings, AoT(One), AoT(Path), AAoT, and MGA exhibit comparable performance, highlighting the AHEAD protocol’s susceptibility. We particularly note that MGA  will be perfectly detected. However, AAoT/AAoG completely evades detection, as show in Section~\ref{exp:tree_detection}). 

The considerable variance in both the attack and LDP outcomes arises from randomly generated queries, whose sizes and baseline responses (i.e., no attacker) can differ substantially. For efficiency, the variance primarily depends on the query’s range size, because wider ranges offer more nodes for fake users to exploit. For response, variance is influenced by both the range size and the real response, producing similar trends in LDP estimation and attacks, as shown in Figure~\ref{fig:tree_differentrho_respond}. 

\textbf{Grid.} Figures~\ref{fig:grid_differentrho_HDG_response} and~\ref{fig:grid_differentrho_HDG_eff} show that our AoG, HAoG, and AAoG achieve high response and efficiency under various parameters, with both AoG and HAoG consistently outperforming MGA. The main distinction between AoG and HAoG lies in whether effective hash pairs are found for each grid; lacking such pairs, AoG naturally degrades to HAoG.

Interestingly, HDG exhibits lower variance in multi-dimensional queries than AHEAD. This is because HDG handles frequencies at a coarser granularity—no more than 64 cells per grid under our default settings—while AHEAD may need to estimate hundreds of nodes. Furthermore, HDG’s response-matrix construction reduces estimation fluctuations, leading to an overall lower variance. 

\foreach \x/\pname in {BlackFriday/BlackFriday}{
    \def\folder{figures/trees/\x/RandomQuery/}
\begin{figure}[hbt!]
  \centering
  \def\subpath{figures/attack/different_eps_and_rho_for_data\pname_rho}
  \def\variable{$\rho$ = }
  \captionsetup[subfigure]{labelformat=parens, labelsep=space}
   \begin{minipage}{0.24\textwidth}
    \includegraphics[width=\linewidth]{\folder \subpath 0.1.png}
    \subcaption{$\rho = 0.1$}
    \label{fig:tree_eps_respond}
  \end{minipage}\begin{minipage}{0.24\textwidth}
    \includegraphics[width=\linewidth]{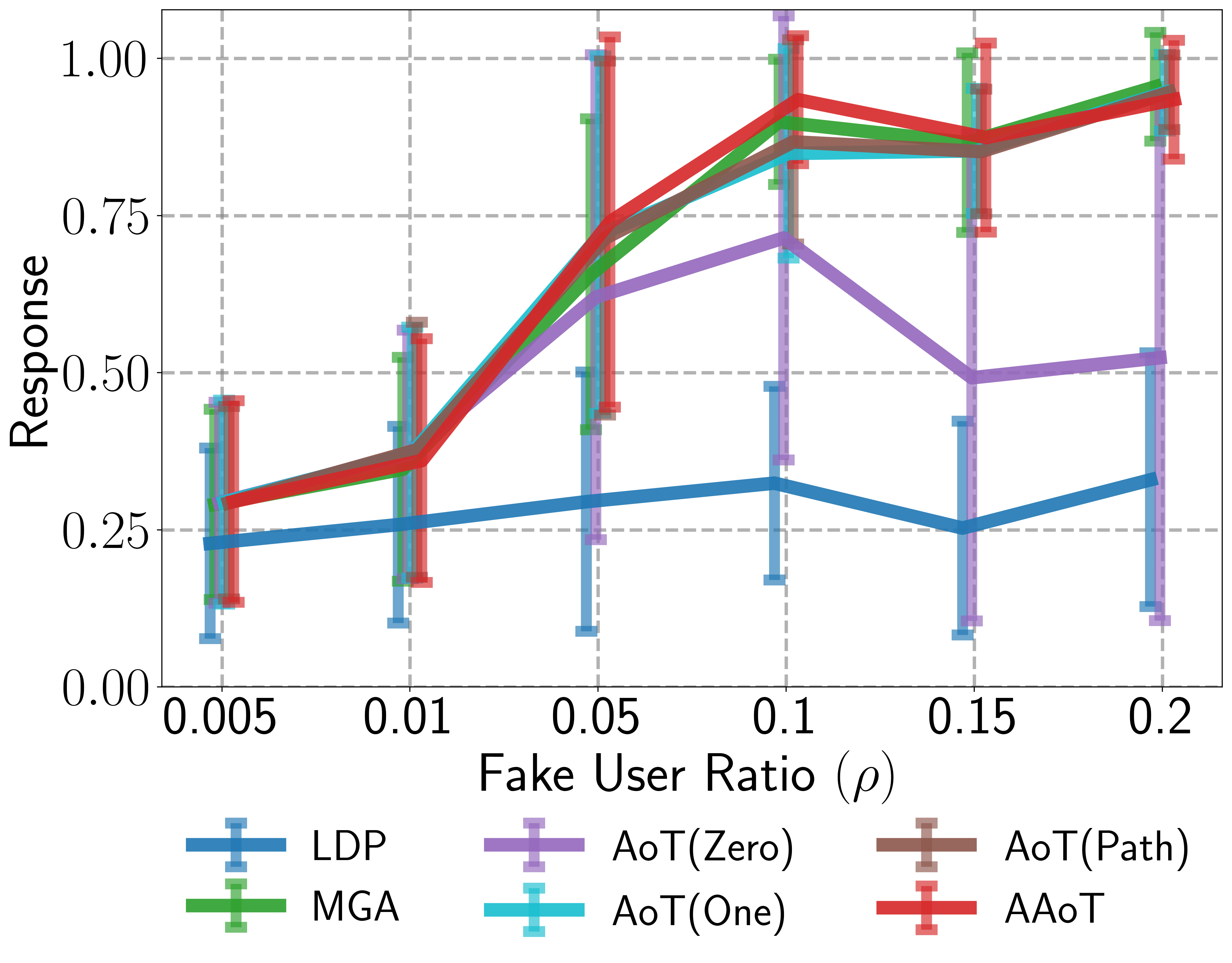}
    \subcaption{$\epsilon = 1.0$}
    \label{fig:tree_rho_respond}
  \end{minipage}
  \caption{The response of AHEAD on BlackFriday dataset of different methods.}
  \label{fig:tree_differentrho_respond}
\end{figure}
\begin{figure}[hbt!]
  \centering
  \def\folder{figures/trees/Efficiency/\x/RandomQuery_littleRho/}
  \def\subpath{figures/attack/different_eps_and_rho_for_data\pname_rho}
  \def\variable{$\rho$ = }
  \captionsetup[subfigure]{labelformat=parens, labelsep=space}
  \begin{minipage}{0.24\textwidth}
    \includegraphics[width=\linewidth]{\folder \subpath 0.1.png}
    \subcaption{$\rho = 0.1$}
    \label{fig:tree_eps_eff}
  \end{minipage}\begin{minipage}{0.24\textwidth}
    \includegraphics[width=\linewidth]{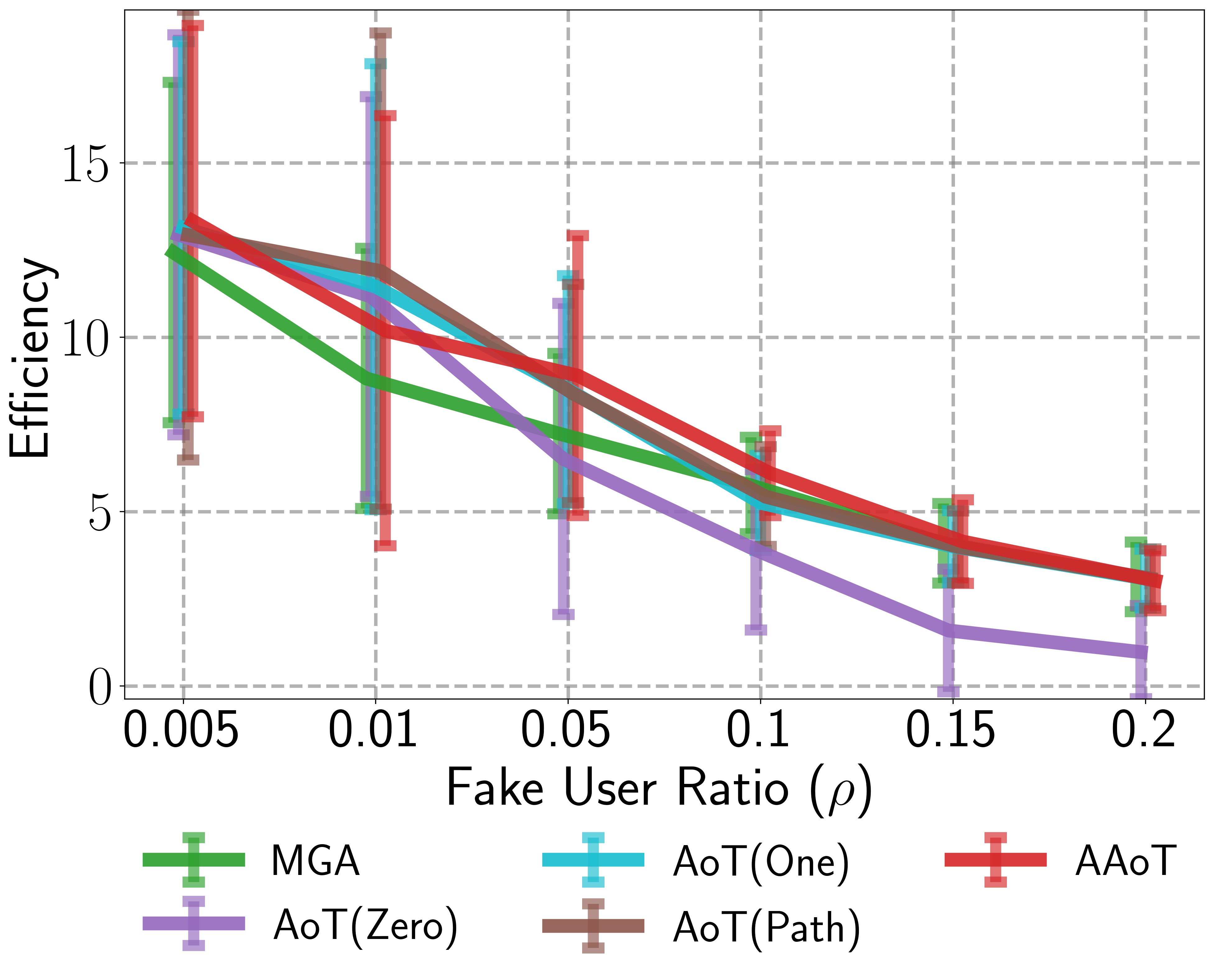}
    \subcaption{$\epsilon = 1.0$}
    \label{fig:tree_rho_eff}
  \end{minipage}
  \caption{The efficiency of AHEAD on BlackFriday dataset of different methods.}
  \label{fig:tree_differentrho_eff}
\end{figure}
}

\begin{figure*}[hbt!]
\def\folder{figures/grid/\x/3-way/}
\def\subpath{attack/different_eps_and_rho_for_data\pname_rho}
\def\variable{$\rho$ = }
\centering
\captionsetup[subfigure]{labelformat=parens, labelsep=space}
\begin{minipage}{0.31\textwidth}
\includegraphics[width=\linewidth]{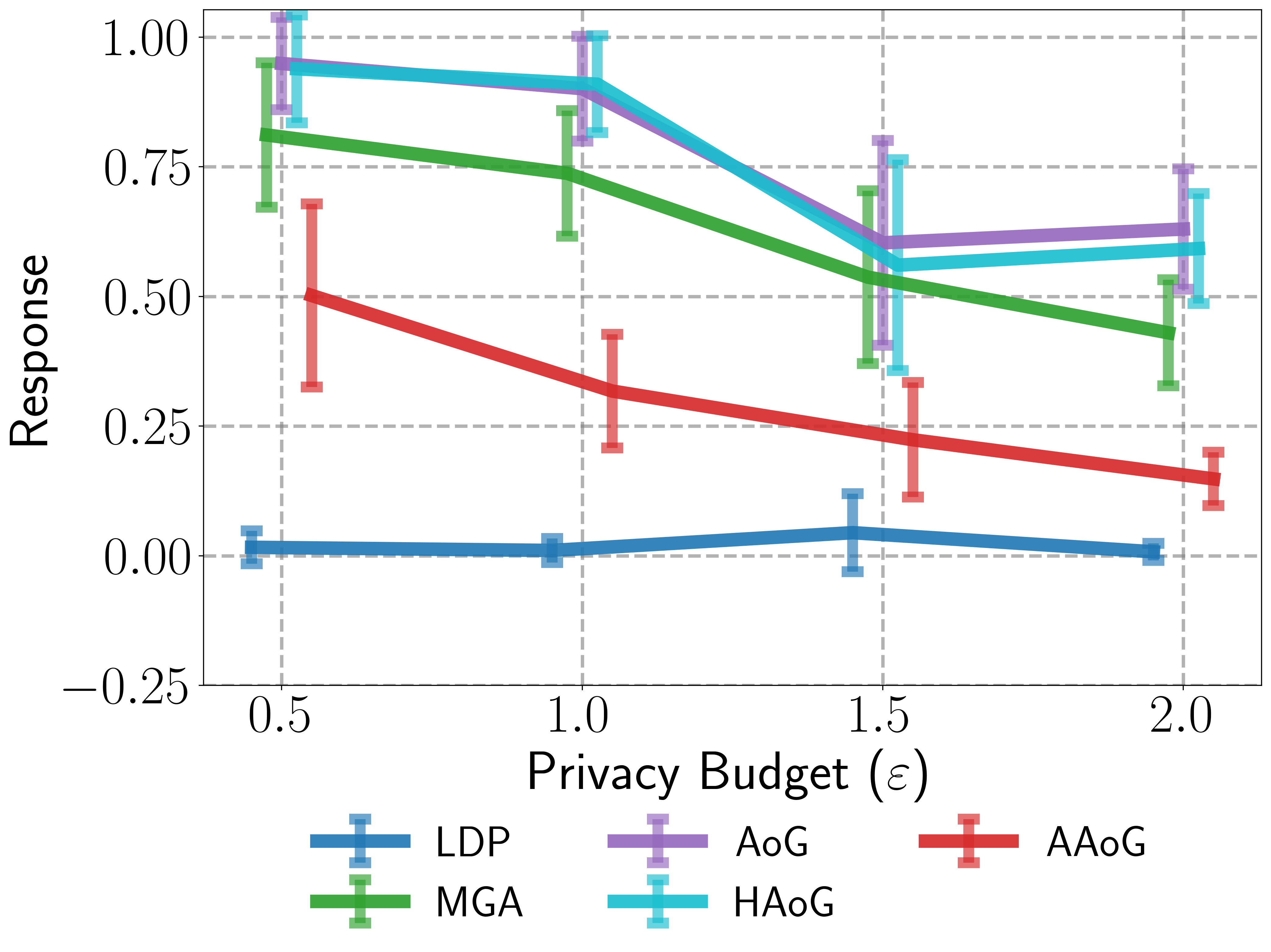}
\subcaption{$\rho = 10\%$, Single Post-process}
\label{fig:grid_eps_respond}
\end{minipage}
\begin{minipage}{0.31\textwidth}
\includegraphics[width=\linewidth]{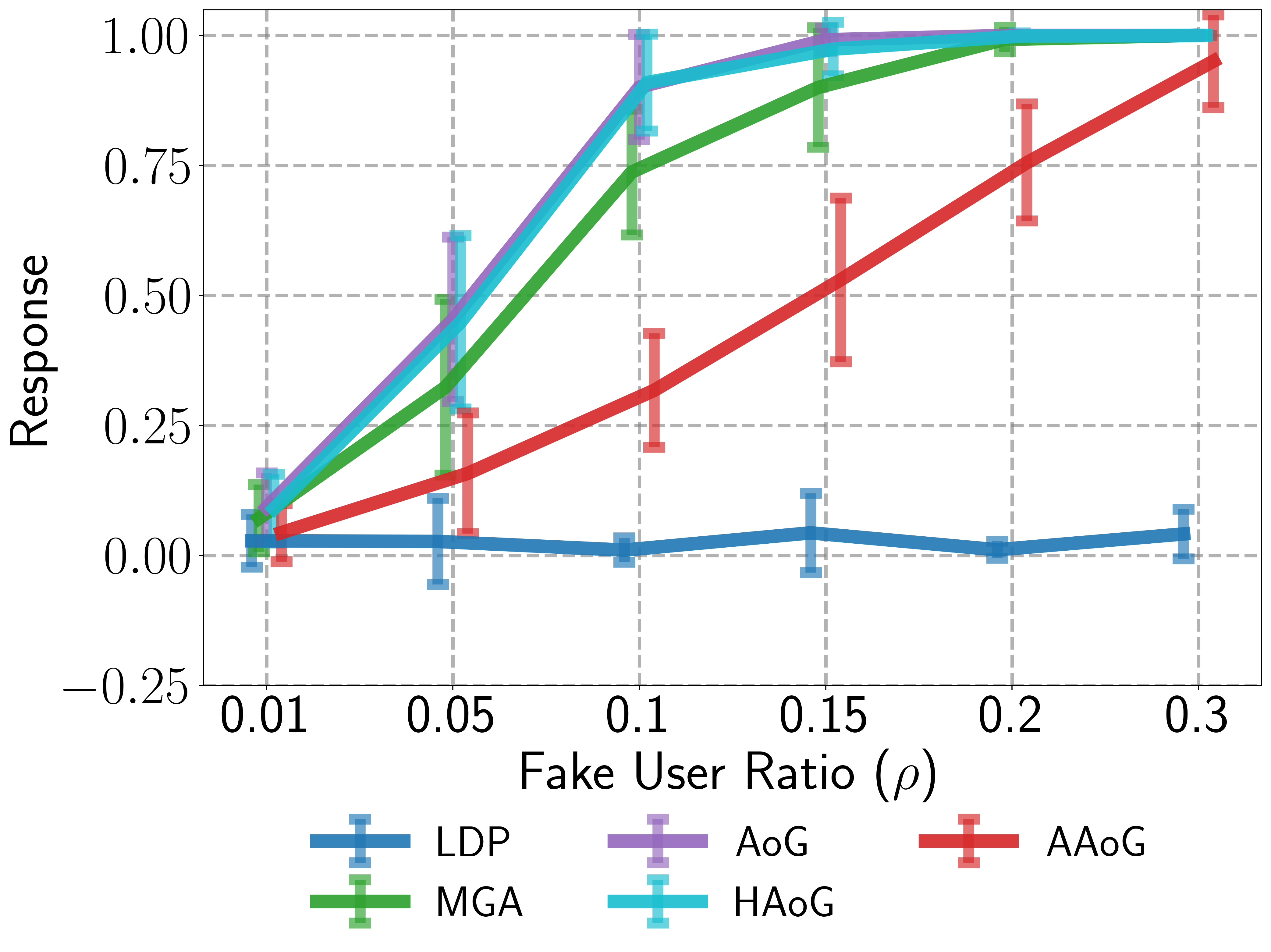}
\subcaption{$\epsilon = 1.0$, Single Post-process}
\label{fig:grid_rho_respond}
\end{minipage}
\begin{minipage}{0.31\textwidth}
\includegraphics[width=\linewidth]{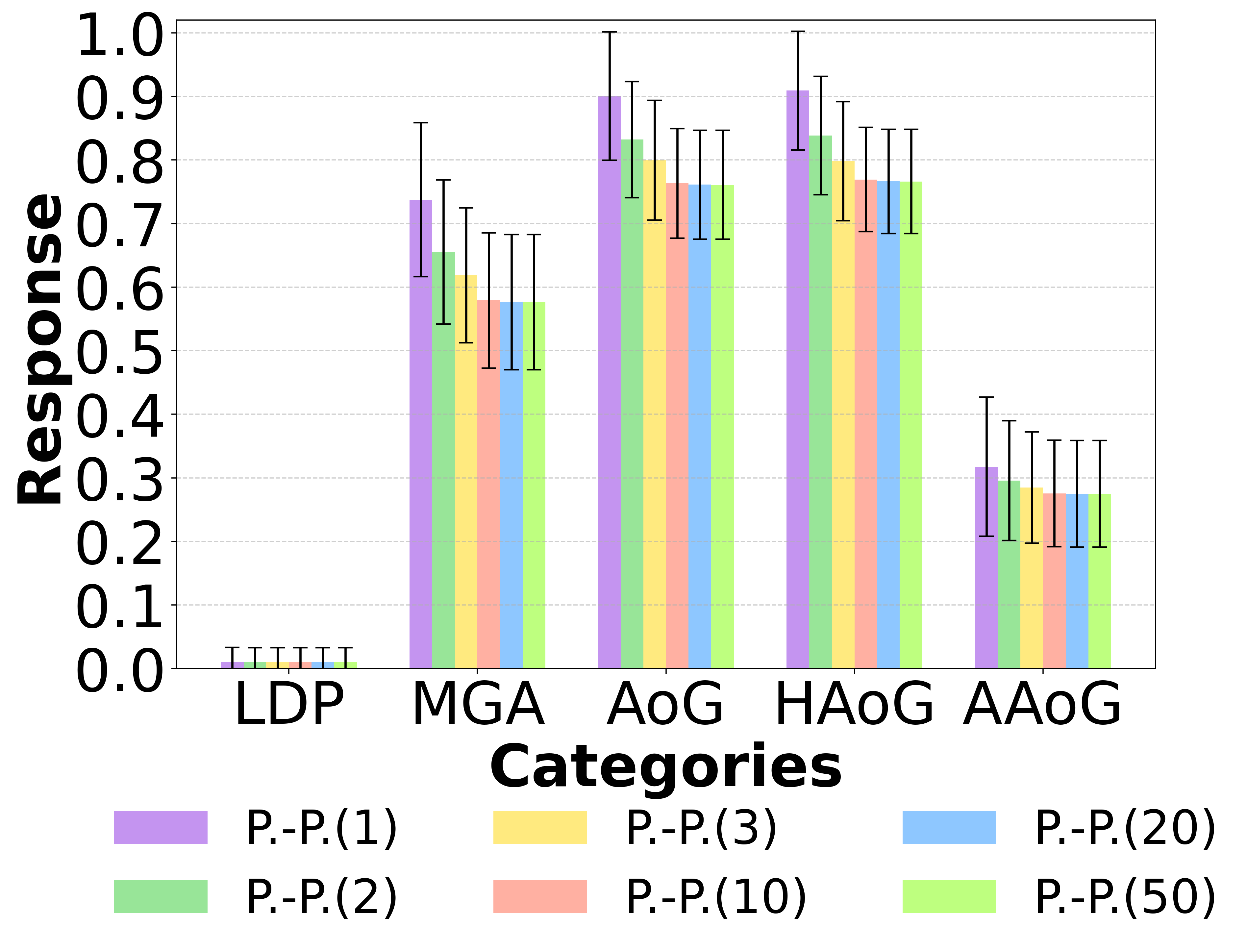}
\subcaption{$\epsilon = 1.0, \rho = 10\%$}
\label{fig:8c}
\end{minipage}
\vspace{-0.3cm}
\caption{The response of HDG on BlackFriday dataset of different methods.}
\label{fig:grid_differentrho_HDG_response}
\end{figure*}
\begin{figure*}[hbt!]
\def\folder{figures/grid/\x/3-way/}
\def\subpath{attack/different_eps_and_rho_for_data\pname_rho}
\def\variable{$\rho$ = }
\centering
\captionsetup[subfigure]{labelformat=parens, labelsep=space}
\begin{minipage}{0.31\textwidth}
\includegraphics[width=\linewidth]{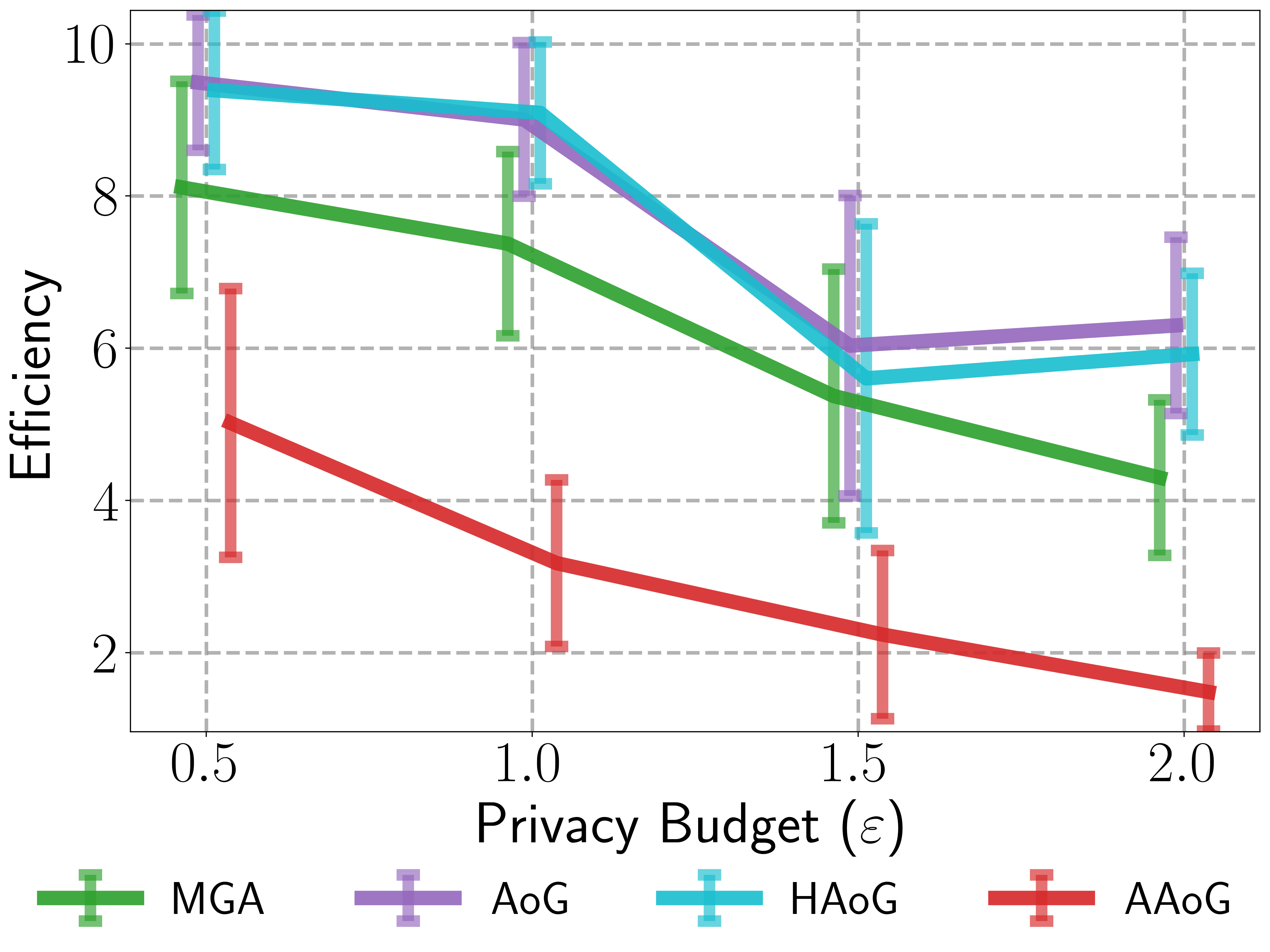}
\subcaption{$\rho = 10\%$, Single Post-process}
\label{fig:grid_eps_eff}
\end{minipage}
\begin{minipage}{0.31\textwidth}
\includegraphics[width=\linewidth]{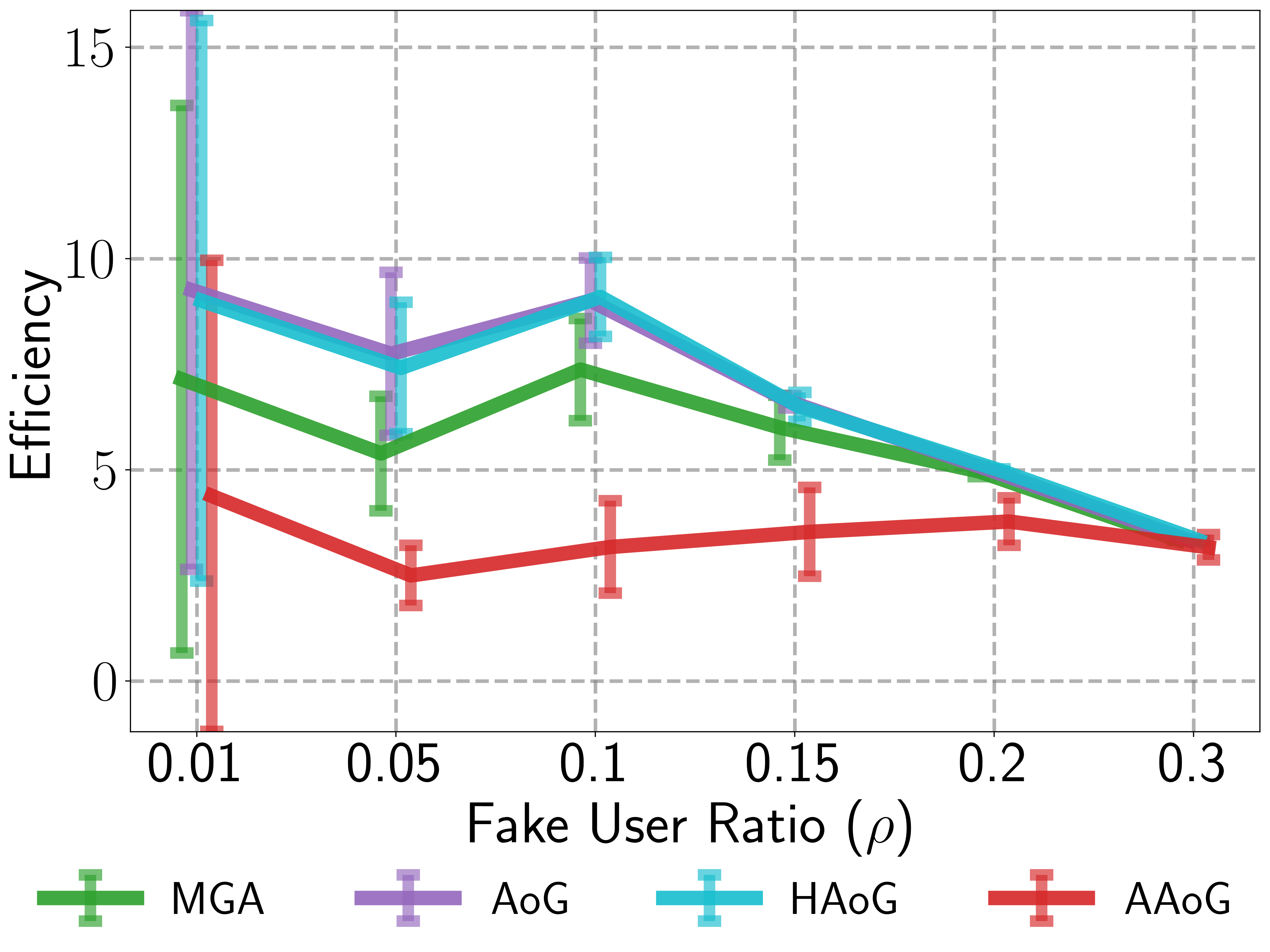}
\subcaption{$\epsilon = 1.0$, Single Post-process}
\label{fig:grid_rho_eff}
\end{minipage}
\begin{minipage}{0.31\textwidth}
\includegraphics[width=\linewidth]{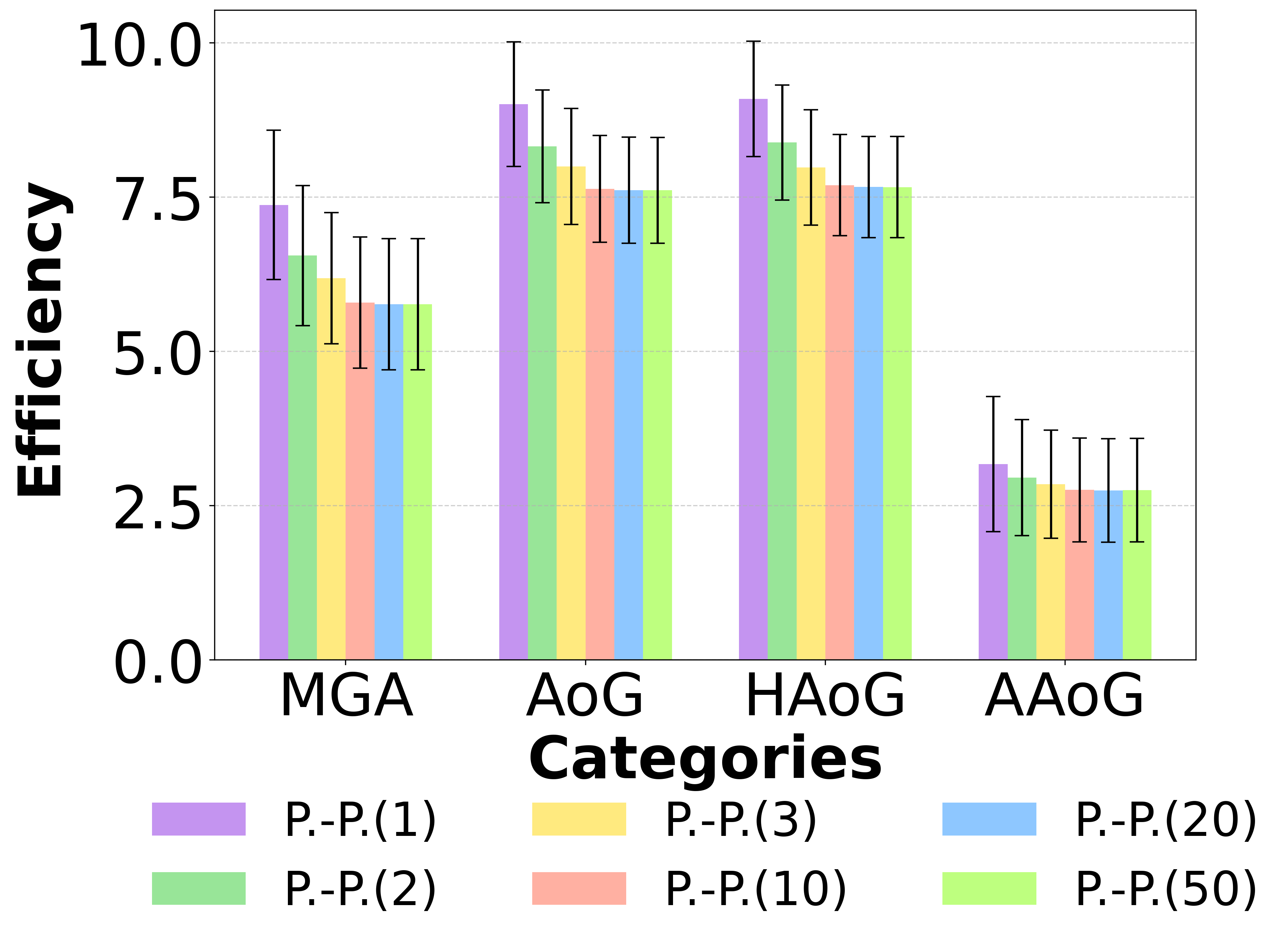}
\subcaption{$\epsilon = 1.0, \rho = 10\%$}
\label{fig:9c}
\end{minipage}
\vspace{-0.2cm}
\caption{The efficiency of HDG on BlackFriday dataset of different methods.}
\label{fig:grid_differentrho_HDG_eff}
\end{figure*}
\subsubsection{Impact of $\epsilon$}\label{sec: Impact of epsilon}
\textbf{Tree.} Figure~\ref{fig:tree_eps_respond} and~\ref{fig:tree_eps_eff} display the effect of varying $\epsilon$. The slight drop in response with increasing \(\epsilon\) stems from the protocol’s improved accuracy in reflecting true frequencies. As \(\epsilon\) grows, both the efficiency and the resulting response decrease accordingly.

In contrast, AoT(Zero) produces weak results because there is insufficient frequency in higher tree layers to trigger further decomposition. Specifically, AHEAD’s adaptive decomposition splits nodes more finely only when their frequency is high. Without enough frequency at upper layers, AoT(Zero) leaves nodes with broad intervals, rendering the optimal assignment strategy inapplicable.

\textbf{Grid.} Figure~\ref{fig:grid_eps_respond} and~\ref{fig:grid_eps_eff} display the effect of varying $\epsilon$. The reduction in response for higher \(\epsilon\) aligns with AHEAD, where HDG gains accuracy as \(\epsilon\) increases. Moreover, OLH uses a larger number of hash keys at higher \(\epsilon\), causing each key to cover fewer cells, which reduces fake-user efficiency for covering the target range. Among our methods, AAoG suffers the most because it spreads its efforts over many hash pairs to avoid detection, unlike AoG and HAoG, which focus on fewer, high-quality pairs. Consequently, AAoG’s response drops more severely as \(\epsilon\) grows, given the diminished effectiveness of each hash pair.

\subsubsection{Impact of $\rho$}\label{sec: Impact of rho}
\textbf{Tree.} Figure~\ref{fig:tree_rho_respond} and~\ref{fig:tree_rho_eff} show the effect of different $\rho$. AoT(One), AoT(Path), and AAoT outperform MGA when $\rho$ is small because our optimal assignment strategy uses fake users more efficiently. When $\rho$ is low, every additional fake user has a greater impact, so a suboptimal approach like MGA incurs a larger performance penalty. However, as $\rho$ grows, the real data’s frequencies are almost entirely dominated by the poisoned data during the Norm-Sub step, rendering the differences between AoT and MGA negligible.

Moreover, efficiency declines as $\rho$ increases due to the Norm-Sub mechanism. As $\rho$ grows, the Norm-Sub threshold $\delta$ rises because of the additional poisoned frequencies. A higher $\delta$ reduces the fraction of frequencies that exceed it, diminishing the effectiveness of injecting more fake frequencies.

\textbf{Grid.} Figure~\ref{fig:grid_rho_respond} and~\ref{fig:grid_rho_eff} display the effect of different $\rho$. Increasing $\rho$ further boosts the response, eventually causing frequencies to fully concentrate in $q_t$ once $\rho$ exceeds 15\%. For AoG, this occurs because, when $\rho$ increases from 10\% to 15\%, $w_2$ decreases from 7 to 5. The lower $w_2$ value relaxes the size constraint, making it easier for AoG to find suitable hash pairs.
\begin{figure}[hbt!]
\centering
\includegraphics[width=\linewidth]{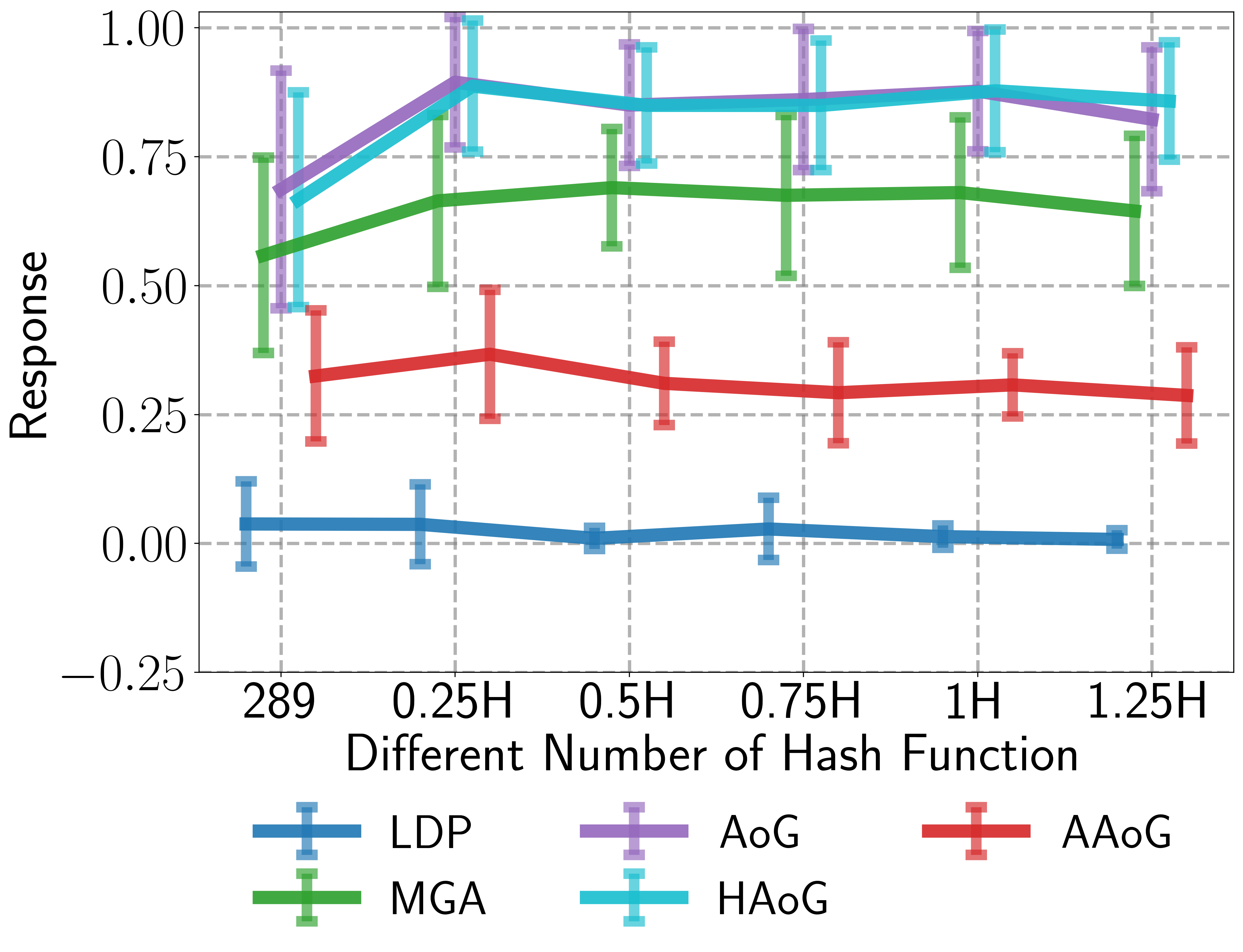}
\vspace{-0.7cm}
\caption{Mismatch in $\mathcal{H}$, with $H = \lfloor\frac{N + M}{15}\rfloor = 11121$}
\label{fig:mismatch_h}
\end{figure}

\begin{figure}[hbt!]
\centering
\begin{minipage}{0.25\textwidth}
\includegraphics[width=\linewidth]{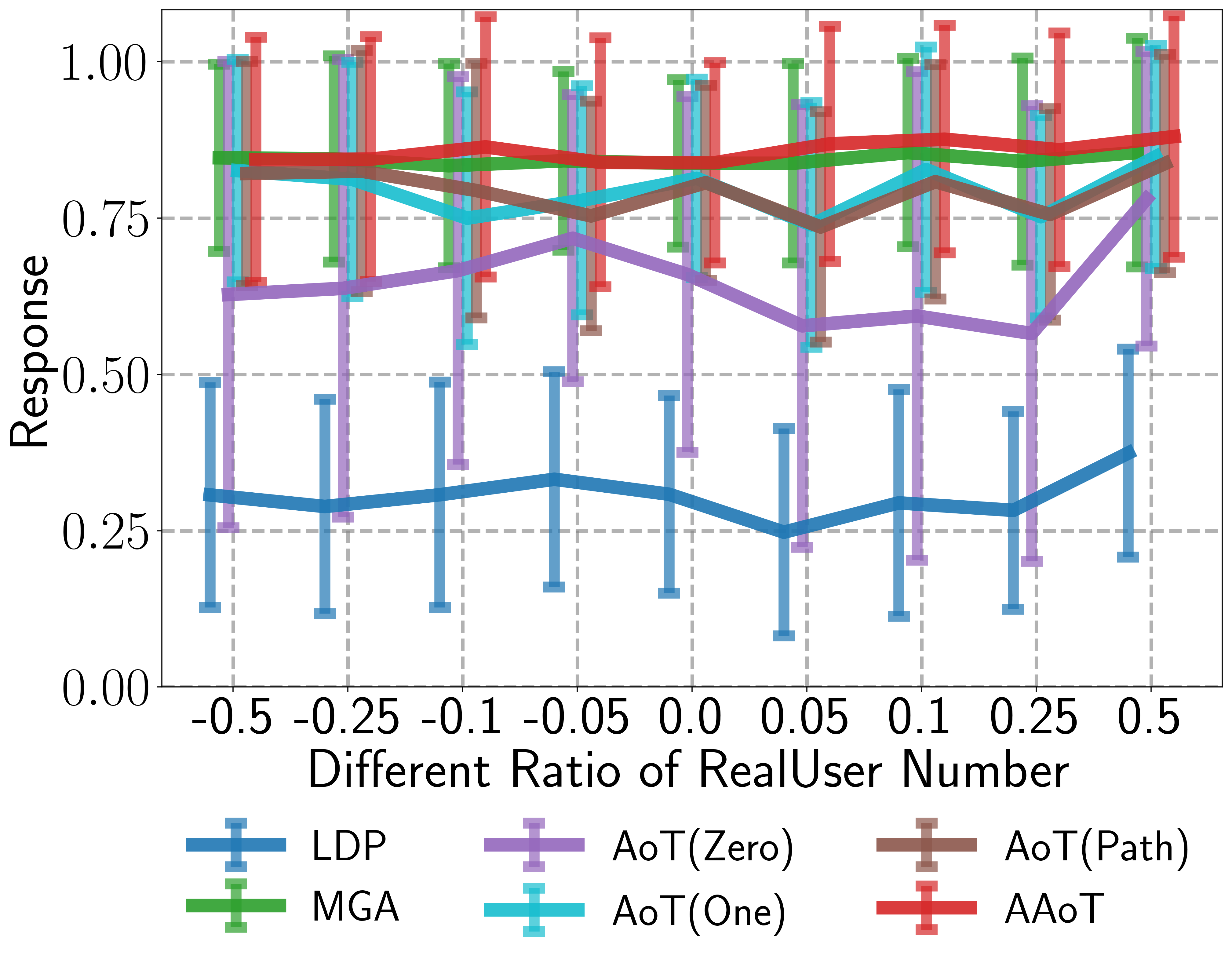}
\subcaption{AHEAD}
\label{fig:mismatch_n_tree}
\end{minipage}\begin{minipage}{0.25\textwidth}
\includegraphics[width=\linewidth]{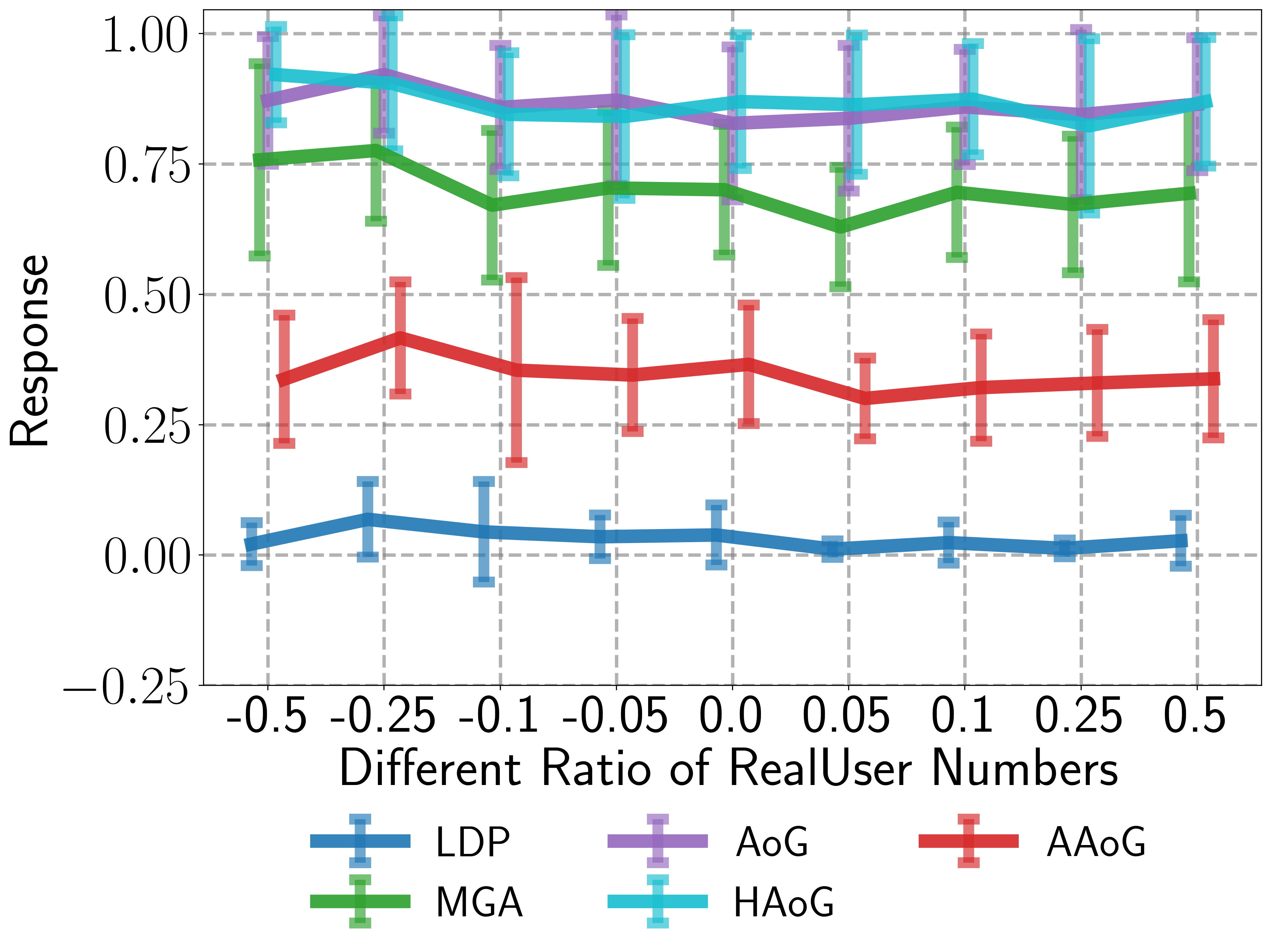}
\subcaption{HDG}
\label{fig:mismatch_n_grid}
\end{minipage}
\vspace{-0.2cm}
\caption{Mismatch on $N$.}
\label{fig:mismatch}
\end{figure}

\begin{figure*}[!hbt]
    \begin{minipage}{0.3\textwidth}
    \includegraphics[width=\linewidth]{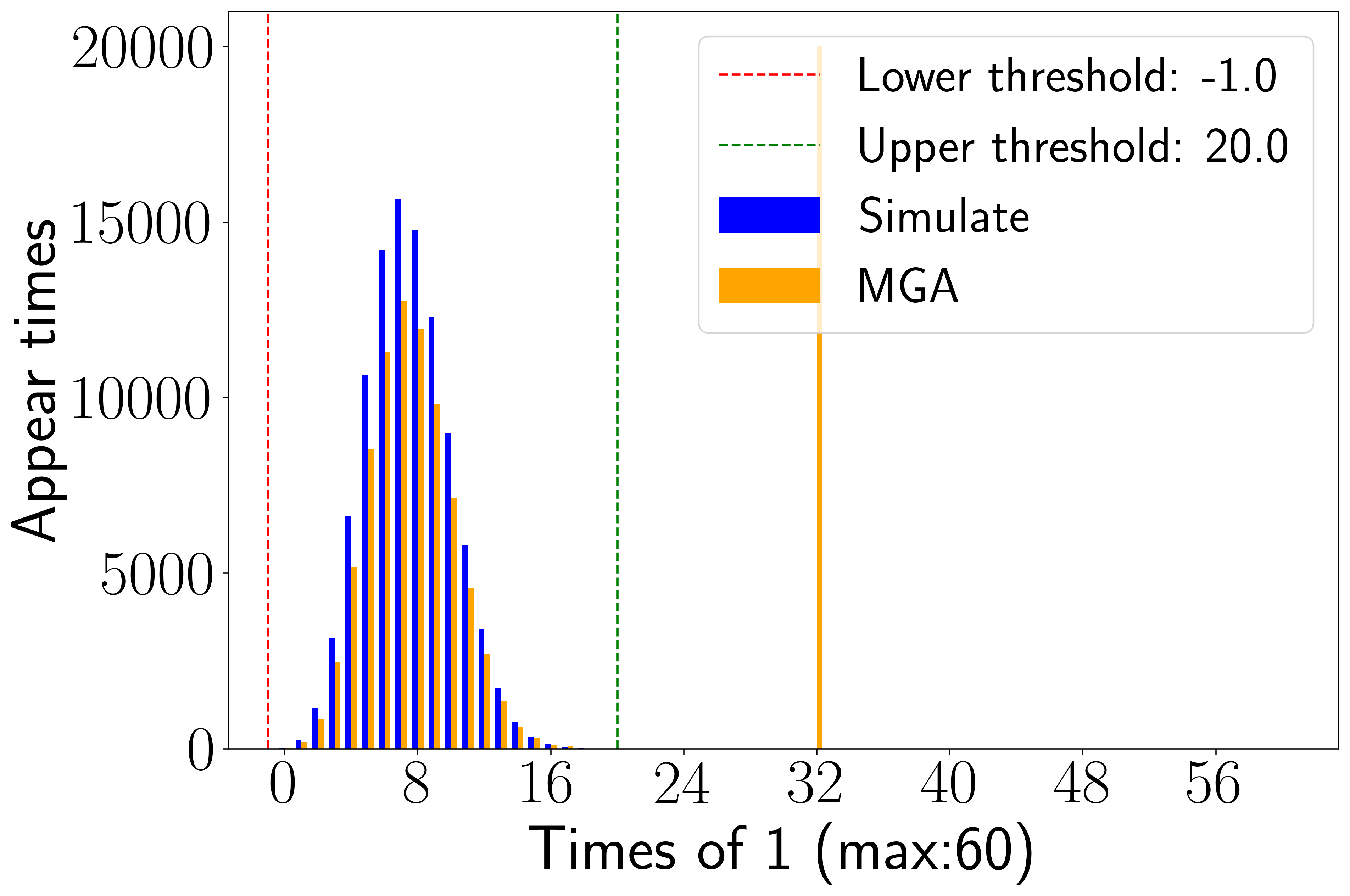}
    \subcaption{MGA}
    \label{fig:mga_1}
  \end{minipage}\begin{minipage}{0.3\textwidth}
    \includegraphics[width=\linewidth]{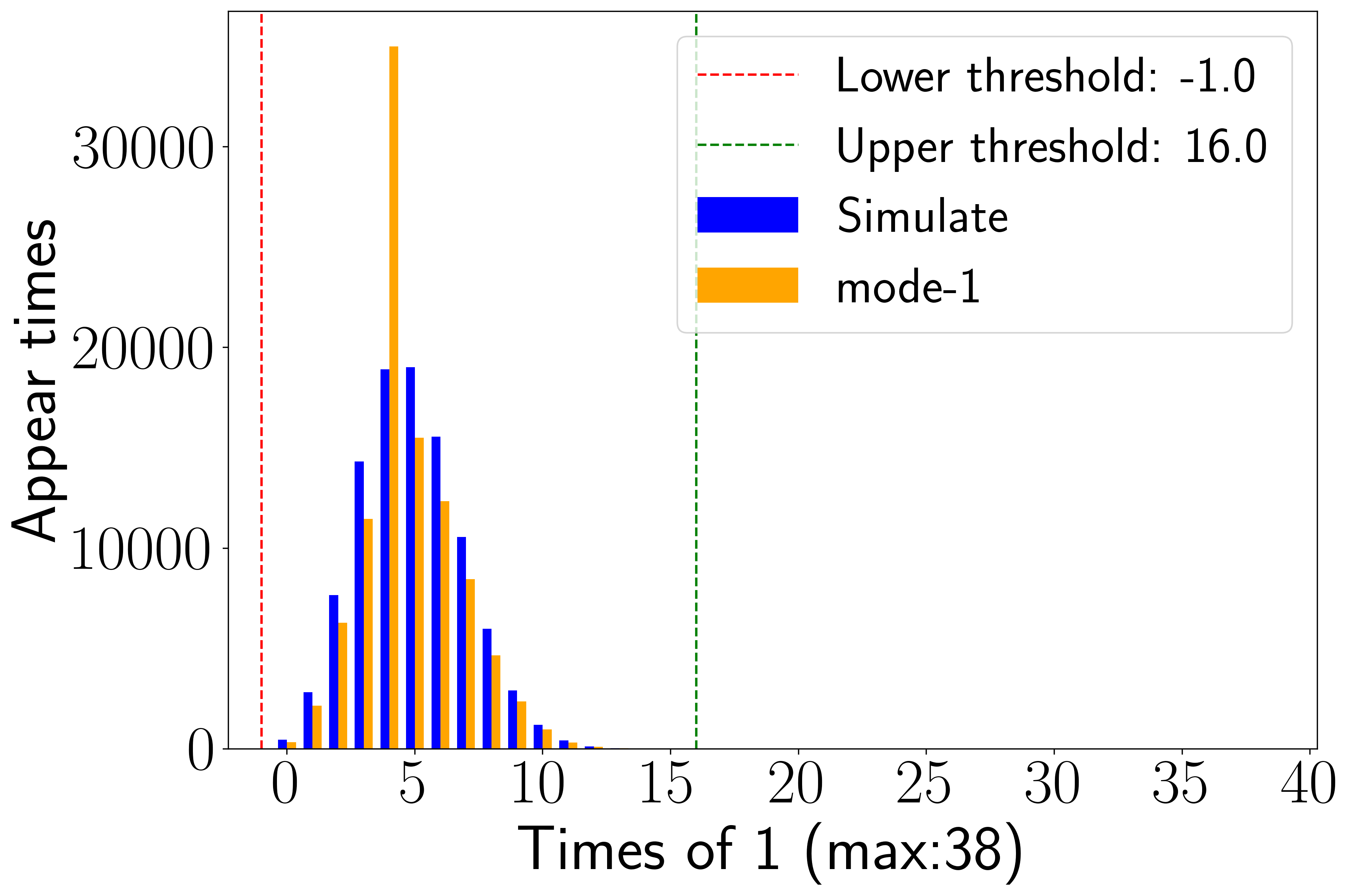}
    \subcaption{AoT(One)}
    \label{fig:aot1}
  \end{minipage}\begin{minipage}{0.3\textwidth}
    \includegraphics[width=\linewidth]{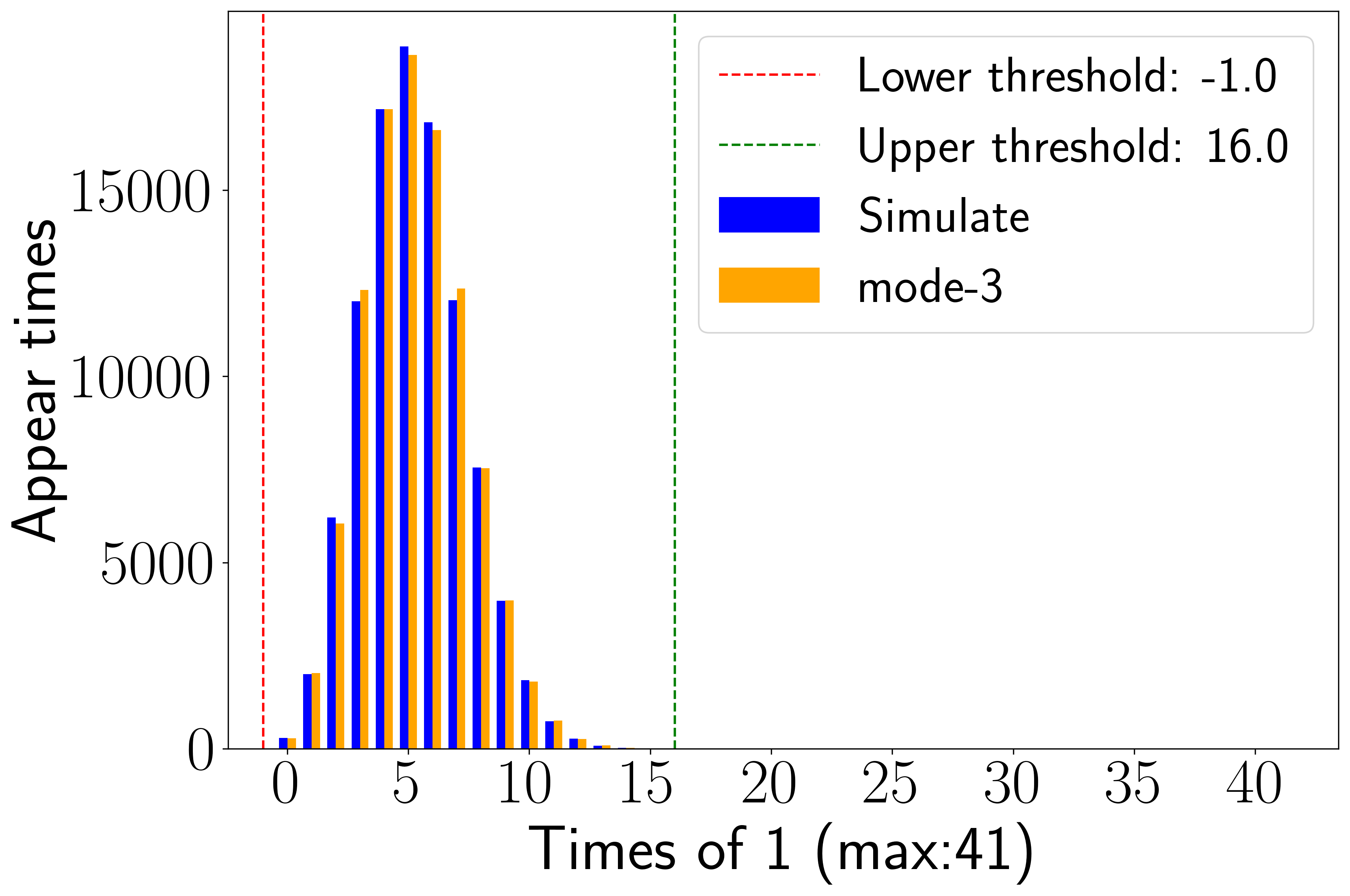}
    \subcaption{AAoT}
    \label{fig:aaot1}
  \end{minipage}
  \vspace{-0.4cm}
  \caption{Demonstration of how the number of 1s distributed when different attack is present.}
\end{figure*}

\begin{figure}[hbt!]
  \centering
  \def\subpath{figures/LDPR/different_eps_and_rho_for_data\pname_rho}
  \def\variable{$\rho$ = }
  \captionsetup[subfigure]{labelformat=parens, labelsep=space}
\begin{minipage}{0.24\textwidth}
    \includegraphics[width=\linewidth]{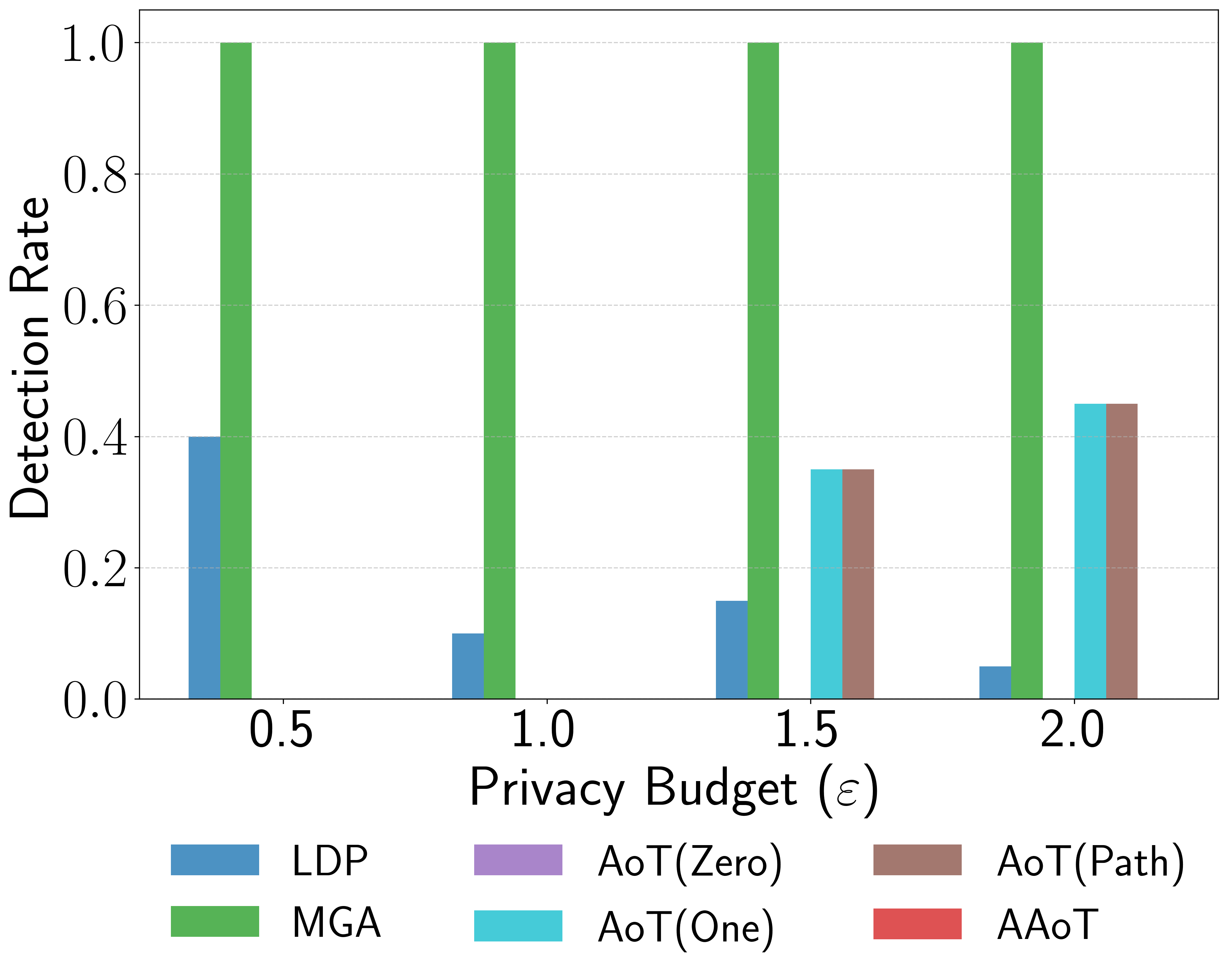}
    \subcaption{Detection Rate for AHEAD}
    \label{fig:tree_detection}
  \end{minipage}\begin{minipage}{0.24\textwidth}
      \includegraphics[width=\linewidth]{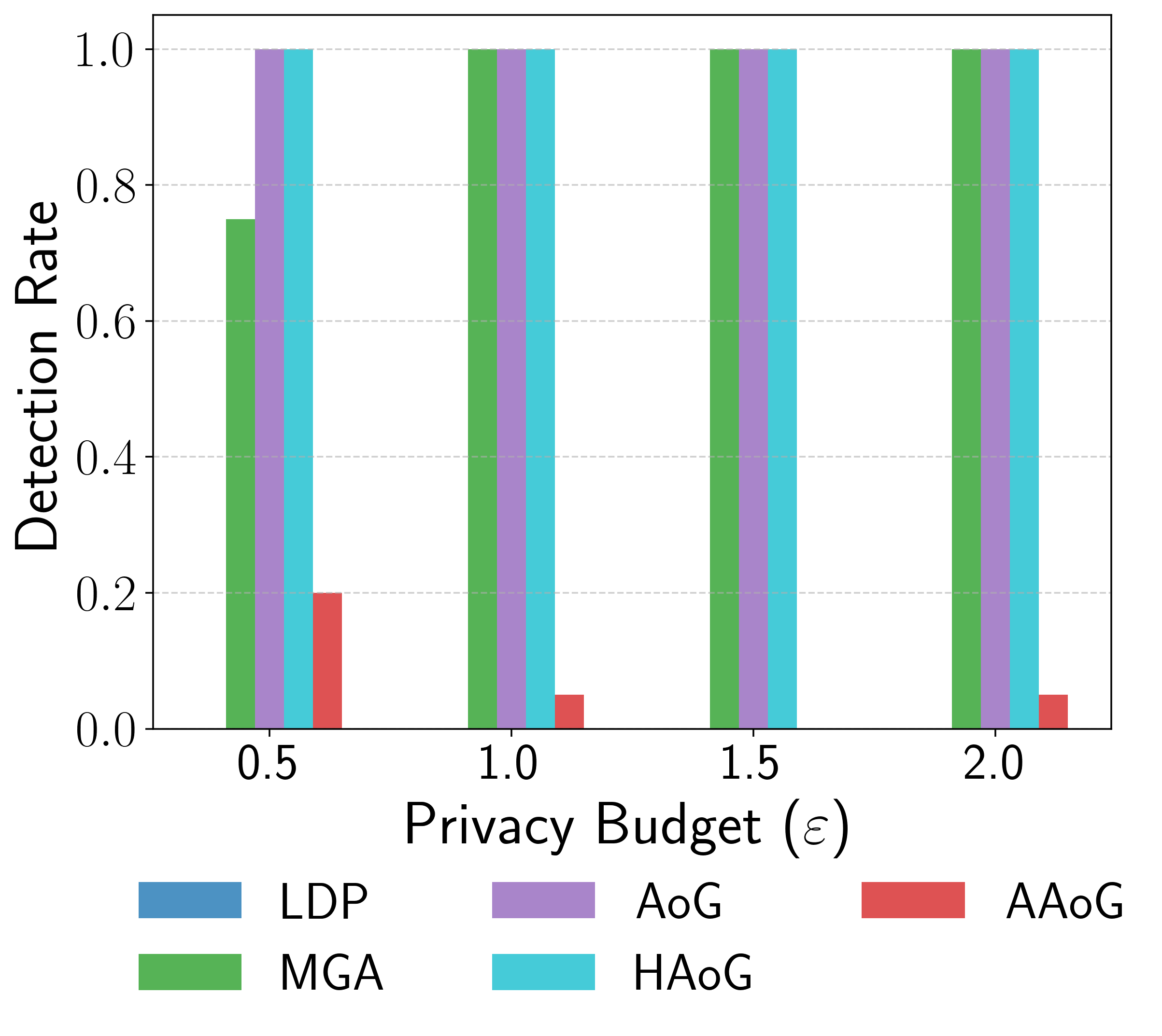}
    \subcaption{Detection Rate for HDG}
    \label{fig:grid_detection}
  \end{minipage}
  \vspace{-0.4cm}
  \caption{Detection result for the attacks.}
  \label{fig:LDPRandDetection}
\end{figure}

\subsubsection{Impact of Number of Rounds in Post-Processing}
Unlike AHEAD, HDG generally conducts multiple post-processing rounds to maintain estimation consistency. Although Theorem~\ref{thm: grid_constraint} formally guarantees optimality for HDG under only a single post-processing round, our attack still proves effective across multiple rounds. The different bar colors in Figure~\ref{fig:8c} and~\ref{fig:9c} illustrate how the response and exploit efficiency evolve under varying numbers of post-processing cycles. Because the consistency step averages cell values inside and outside $q_t$, each additional round further reduces the ultimate response. Even so, the persistently high final response confirms that most of the damage inflicted during the first-round Norm-Sub remains largely irreversible.

\subsubsection{Impact of Mismatching Victim Protocols}\label{exp:other_protocol}
In Figure A.32 in the Appendix, we present our attack’s performance on several additional LDP range query protocols. Although applying the attack requires protocol-specific adaptations—for example, modifying the algorithms or adjusting the protocols themselves—the core approach remains unchanged. Specifically, for privNUD~\cite{privnud}, we implement AoT in a layer-wise fashion rather than relying on the protocol’s dynamic allocation. For PriPL~\cite{pripl}, we simplify the PL fitting process because the SW mechanism~\cite{sw} replaces OUE, rendering AoT’s usual approach inapplicable. Finally, when attacking DHT~\cite{dht}, our strategy closely mirrors MGA, as the optimal assignment effectively involves greedily injecting frequencies into the target range.

Figure A.32 in the Appendix illustrates our attack’s adaptability across these protocols. AoT works for PrivNUD nearly as effectively as in AHEAD because both rely on similar post-processing, highlighting Norm-Sub’s vulnerability. PriPL, however, achieves slightly weaker outcomes because it employs a normalization process akin to Norm-Sub but also enforces parent-child consistency, reducing the extent to which fake frequencies can be amplified compared to AHEAD, HDG, or PrivNUD. Lastly, DHT sees the least pronounced effect; each user submits only a single bit of information, sharply constraining fake-user efficiency. Furthermore, DHT lacks the Norm-Sub normalization process, reinforcing our claim that Norm-Sub inherently magnifies the impact of data poisoning.

\subsubsection{Impact of Mismatching Hash Functions}

Figure~\ref{fig:mismatch_h} displays the results obtained when the server varies the number of hash functions. Because the server can select any number of hash functions for OLH, as long as they belong to a universal hash family, we set a minimum of $17^2 = 289$\footnote{Here, 17 is a prime greater than the number of cells in any grid, enabling a linear congruential generator~\cite{universalhash} with 289 hash functions. Depending on $g_1$ and $g_2$, the minimum number of hash functions in our experiment can be either $17^2=289$ or $37^2 = 1369$}. While our attack is indeed affected by the choice of $|\mathcal{H}|$, the impact only becomes significant when $|\mathcal{H}|$ is extremely small. Experimental results on other dataset can be found in Figure G.60 in the Appendix.

\subsubsection{Impact of Mismatching Distributions}
In Figures A.30 and G.58 in the Appendix, we examine how queries with either large or small real responses are affected. We simulate real data using synthetic datasets by randomly selecting range centers that are either close to or far from the mean. Centers with a distance to the data mean smaller or larger than one standard deviation are classified as matching or mismatching the data distribution, respectively. The results show that our attack’s performance does not hinge on the magnitude of the real response. However, allocating a high $\rho$ to queries with large real responses may reduce efficiency, since frequencies cannot exceed 1.

\subsubsection{Impact of Mismatching User Count}\label{exp:treemismatchN}
Figure~\ref{fig:mismatch_n_tree} and Figure~\ref{fig:mismatch_n_grid} illustrate the impact of attackers misestimating the number of real users. Each data point has a fixed \(\rho\), but the assumed number of real users varies. The results suggest that the attacks do not critically depend on knowing the precise value of $N$. Although AoT uses $N$ to approximate frequencies, any performance degradation from misestimating \(N\) appears minor. Additional experimental findings on other datasets can be found in Figures A.31 and G.59 in the Appendix. 
\subsubsection{Detection}\label{exp:tree_detection}
\textbf{Tree.} In Figure~\ref{fig:tree_detection}, the detection mechanism consistently flags MGA and occasionally AoT. By contrast, AAoT evades detection almost entirely, mirroring the low detection rate of the no-attack (LDP) scenario (although the figure highlights an outlier, on average both AAoT and LDP have around a 5\% detection rate; see Section~\ref{exp:tree_detection_more}). MGA’s vulnerability stems from its indiscriminate reporting of 1s via OUE for all nodes in the target range, while our optimal assignment for AoT adds only the necessary number of 1s. As a result, AoT achieves both optimality and greater stealth, evading detection in roughly half the cases.

By modeling the genuine distribution of 1s, AAoT becomes virtually indistinguishable from legitimate users, even during its poisoning activities. As shown in Figures~\ref{fig:mga_1}, \ref{fig:aot1}, and \ref{fig:aaot1}, the histograms of 1s for MGA and AoT exhibit anomalous patterns—leading to MGA’s consistent detection—whereas AAoT’s histogram aligns with typical user data, avoiding statistical flags.

\textbf{Grid.} Figure~\ref{fig:grid_detection} shows the detection result on HDG. The detection mechanism readily identifies AoG, HAoG, and MGA in grid-based settings, which is simliar to AHEAD. When no fake users are present, the detector triggers only sporadically (see the LDP bar in Figure~\ref{fig:grid_detection}). However, our AAoG strategy remains undetected by opting for a large number of hash functions. By allowing a 10\% detection tolerance—meaning there is under a 10\% chance of being caught—AAoG evades detection in the majority of trials.

\section{Conclusion} \label{sec: conclusion}
We present a comprehensive study of data poisoning attacks on LDP range query protocols. Our AoT (tree-based) and AoG (grid-based) attacks achieve provably optimal effectiveness by leveraging Norm-Sub post-processing to magnify impact. Through extensive theoretical analysis and empirical evaluation, we show they can significantly skew range query estimations, offering 5–10x more influence than a normal user with only a small fraction of malicious users. We propose a detection-based defense and adaptive attacks that maintain effectiveness while evading detection.

Our findings expose critical vulnerabilities in current LDP range query protocols, underscoring the urgent need for robust defenses in real-world applications. Future research includes designing more resilient protocols and exploring defense strategies that balance utility with security against advanced poisoning attacks.

\clearpage
\newpage

\section{Ethics Considerations}
The ethical considerations for this research were carefully reviewed in accordance with the USENIX Security '25 guidelines. While this paper introduces novel attack strategies and adaptive mechanisms targeting vulnerabilities in local differential privacy (LDP) protocols, the research was conducted with the primary goal of advancing the understanding of these vulnerabilities and improving the security of existing systems. The attacks were evaluated solely in controlled environments using synthetic and publicly available datasets, ensuring no harm to real-world systems or users. We explicitly state that these methods will not be deployed against any entity, nor will we encourage or authorize others to do so. Instead, the findings aim to stimulate constructive discussions within the research community and drive the development of more robust LDP protocols. Furthermore, we proposed detection-based defenses and adaptive countermeasures to mitigate the risks posed by such attacks, emphasizing the importance of balancing utility and security. This work adheres to ethical principles by prioritizing transparency, minimizing potential harm, and contributing to the broader goal of improving privacy-preserving technologies.

\section{Open Science}
To comply with the USENIX Security open science policy, we have adhered to the principles of transparency and reproducibility by openly sharing the research artifacts associated with this paper. Specifically, we will provide access to datasets, scripts, and source code used in our experiments, ensuring that our findings can be independently verified and replicated. These artifacts are available for the Artifact Evaluation committee and will remain accessible after publication. In cases where certain artifacts cannot be shared due to licensing restrictions or privacy concerns, we have provided detailed justifications in the supplementary materials. This commitment aligns with the broader goal of fostering open science and advancing the reproducibility of scientific research.

\clearpage
\newpage

{\footnotesize \bibliographystyle{acm}
\bibliography{Reference}}

\clearpage
\newpage

\appendix
\section{Dataset Description}\label{sec: Dataset Description}
The description of five real-world datasets (BFive~\cite{bfive}, BlackFriday~\cite{BlackFriday}, Loan~\cite{loan}, PaySim~\cite{paysim}, Salaries~\cite{salaries}) used in our experiments is shown below. 
\begin{itemize}
    \item \textbf{Synthetic}: Contains 1 million synthetic data randomly sampling Gaussian and Laplace distribution. The entire domainis $[0,1023]$, aligning with AHEAD's domain. The distribution has mean 512 and standard deviation 40.
    
    \item \textbf{BFive}~\cite{bfive}: Contains approximately 1 million records from an interactive online personality test, where each entry captures the time spent on individual questions.
    
    \item \textbf{BlackFriday}~\cite{BlackFriday}: Comprises 500 thousand sales records from Black Friday events, including both buyer information and purchased items. This dataset aligns with the incentive scenario described in Section~\ref{sec: Introduction}, illustrating how fake purchase demands might be created.
    
    \item \textbf{Loan}~\cite{loan}: Includes 2.2 million records on accepted and rejected loans; we sampled 1 million of these for our experiments.
    
    \item \textbf{PaySim}~\cite{paysim}: Consists of synthetic data generated to simulate a private financial dataset, mirroring realistic transaction patterns. We sampled 1 million records out of the original 6 million.
    
    \item \textbf{Salaries}~\cite{salaries}: Contains around 14 thousand records detailing the names, job titles, and compensation of San Francisco city employees from 2011 to 2014.
\end{itemize}
\section{Notation Table}
We summarize notations used in this paper in Table~\ref{table: Notation Table}.

\begin{table}[t!]
\centering
\begin{tabular}{@{}|p{2.2cm}||p{5.5cm}|@{}}
\hline
&Description \\
\hline
$\epsilon$ & Privacy Budget. \\
\hline
$[d]$ & $[0,\dots,d-1]$. The interval from $0$ to $d-1$. \\
\hline
$p,q$ & $\frac{1}{2}$ and $\frac{1}{e^\epsilon + 1}$ respectively. \\
\hline
$g$ & $e^\epsilon + 1$ rounding to the nearest integer. The number of hash key in OLH.\\
\hline
$d$ & The number of dimension. \\
\hline
$q_t$ & The target query that attacker is trying to maximize.\\
\hline
$X_{fake}, X_{real}$ & Data sent by fake users and real users respectively after local perturbation.\\
\hline
$\mathcal{M}_{X_{\text{real}} \cup X_{\text{fake}}}(q_t)$ & The response from server who estimates the range query $q_t$ using data from $X_{\text{real}}$ and $ X_{\text{fake}}$. \\
\hline
$N,M$ & The number of real, fake users respectively. \\
\hline
$\rho$ & The ratio of fake users over all users. $\frac{M}{N+M}$. \\
\hline
$\mathcal{H}, \mathcal{C}$ & The set of hash functions and hash keys. $\mathcal{C} = [g]$. \\
\hline
$g_1, g_2$ & The number of cells in 1-D grids and the number of columns (rows) in 2-D grids in HDG. \\
\hline
$\mathcal{G}_1$ & the set of 1-D cells. $[g_1]$ \\
\hline
$\mathcal{G}_2$ & the set of 2-D cells. $[g_2] \times [g_2]$. \\
\hline
$R_{q_t}(G)$ & The set of cells in grid $G$ that are inside the range of range query $q_t$.\\
\hline
$\delta$ & The threshold in Norm-Sub operation.\\
\hline
$\beta$ & The probability of failing an attack. \\ 
\hline
$\alpha$ & The significant threshold of server. \\
\hline
\end{tabular}\vspace{0.5cm}
\caption{Notation Table}
\label{table: Notation Table}
\end{table}

\section{The Missing Psuedocode}\label{sec: The Missing Psuedocode}
\subsection{Strawman Method: MGA}\label{sec: Strawman Method: MGA}
In Section \ref{sec:baseline}, we introduce two attacks that are modified from MGA in frequency estimation. Here are the detailed procedures.
\begin{algorithm}
    \DontPrintSemicolon
    \caption{Baseline: MGA on Tree}\label{algo: MGA_on_Tree}
    \textbf{Parameter}: $p,q$ (the probabilities used in OUE). $q_t$ (target query). \\
    \While{Server requests layer $L$}{
        \For{user $u_i$ on layer $L$}{
            $k \leftarrow $ number of nodes inside $q_t$ in $L$\\
            $S \leftarrow$ randomly draw $\max(\lfloor p + (|l|-1)q - k \rfloor,0)$ nodes from outside $q_t$ in $L$\\
            user $u_i$ submits $1$ for nodes inside $q_t$ and nodes in $S$, and $0$ otherwise.\\
        }
    }
\end{algorithm}

\begin{algorithm}
    \DontPrintSemicolon
    \caption{Baseline: MGA on Grid}\label{algo: MGA_on_Grid}
    \textbf{Parameter}: $q_t$ (target query). $\mathcal{H}$ (the set of hash functions). $\mathcal{C} = [g]$ (the set of hash keys). $\mathcal{G}_1 = [g_1]$ (the set of cells in 1-D grids). $\mathcal{G}_2 = [g_2^2]$ (the set of cells in 2-D grids).

    $S_1[h,c] = \{v \mid h(v) = c, v \in \mathcal{G}_1\} \ \forall h \in \mathcal{H}, c \in \mathcal{C}$. \\
    $S_2[h,c] = \{v \mid h(v) = c, v \in \mathcal{G}_2\} \ \forall h \in \mathcal{H}, c \in \mathcal{C}$.

    \For{every $x$-D grid $G_i$}{
        $R_{q_t}(G_i) \leftarrow$ the set of cells in $G_i$ that lie within $q_t$. \\
        every user $u$ in $G_i$ submits $(h,c) = \arg\max_{(h,c)\in \mathcal{H}\times\mathcal{C}} |S_x[h,c] \cap R_{q_t}(G_i)|$,
        breaking ties randomly. \\
    }
\end{algorithm}

\subsection{Main Algorithms}\label{sec: Main Algorithms}
Algorithms~\ref{algo: AoT},~\ref{algo: AoG}, and~\ref{algo: AoG_heuristic} are the pseudocode for AoT, AoG, and HAoG, respectively. 

\begin{algorithm}[t!] 
    \DontPrintSemicolon
    \caption{Attack on Tree (AoT)}\label{algo: AoT}
    \textbf{Parameter}: $q_t$ (target query). $s$ (heuristic strategy when all tree coefficient are $0$, can either be Zero, One or Path). \\
    \SetKwProg{Fn}{Function}{}{}
    \SetKwFunction{getCoef}{getCoef}
    \Fn{\getCoef{$T,q$}:\tcp*[f]{Calculate $c_v^{q_t,T}$}}{\label{algo: AoT_coef_start} %
        $C[v] \leftarrow 0 $ for all $v \in T$ \\
        \SetKwFunction{downward}{downward}
        \Fn{\downward{$v, c$}:}{
            \If{$v$.child $= \emptyset$}{
                $C[v] \leftarrow C[v] + c$ \\
                \textbf{Return}
            }
            $\lambda \leftarrow \frac{|v.child|}{|v.child| + 1}$\\
            $C[v] \leftarrow C[v] + \lambda c$\\
            \For{$child \in v$.child}{
                \downward{$child, (1-\lambda)c$}\\
            }
        }
        $V \leftarrow$ Nodes in $T$ that are used to estimate $q_t$.\\
        \For{$v \in V$}{
            \downward{$v,1$}\\
        }
        \textbf{Return} $C$
    }
    \label{algo: AoT_coef_end}
    \While{Server requests for layer $L$}{
        updateTree($L$) \tcp{Update tree struct. by adding $L$}
        \label{algo: aot_getTree}$T \leftarrow$ getTree() \tcp{Generate the tree structure.}
        \label{algo: aot_getCoef}$C \leftarrow$ \getCoef{$T,q_t$} \\
        \If{$C$ is all $0$ in $L$}{
            Fake users return according to $s$. \\
            \textbf{Continue}
        }
        $L_{sorted} \leftarrow $Sort $v \in L$ using $C[v]$ as key\\
        $M_L \leftarrow$ The number of fake users at $L$\\
        $A \leftarrow \{<\underbrace{M_L,M_L,...,M_L}_{k\mbox{ }times},i,0,...,0>\}$ for all $0 \leq k \leq |L|, 0 \leq i < M_L$ \tcp{Enumerate all potential optimal assignment}\label{algo: AoT_all_ass}
        Calculate $\hat{f}_a(v)$ for $v\in L$ and $a\in A$ by performing OLH and Norm-Sub, assuming uniformly distributed real user data. \\
        $a_{best} \leftarrow \max_{a \in A} \sum_{v \in L} C[v]\hat{f}_a(v)$ \\\label{algo: aot_best}
        Fake users return according to $a_{best}$ and $L_{sorted}$.\label{algo: aot_return}
    }
\end{algorithm}

\begin{algorithm}[t!]
    \DontPrintSemicolon
    \caption{Attack on Grid (AoG)}
    \label{algo: AoG}
    \textbf{Parameter}: 
    \(\mathcal{H}\) (set of hash functions), 
    \(\mathcal{C} = [g]\) (set of hash keys), 
    \(w_1, w_2\) (Size constraints from Theorem~\ref{thm: grid_constraint}), 
    \(A_{q_t}\) (set of attributes in \(q_t\)), 
    \(\mathcal{G}_1 = [g_1]\) (cells in 1-D grids), 
    \(\mathcal{G}_2 = [g_2^2]\) (cells in 2-D grids).

    \SetKwProg{Fn}{Function}{}{}
    \SetKwFunction{columnconstraint}{columnConstraint}
    \SetKwFunction{findcover}{findHashPair}

    \Fn{\columnconstraint{$S, G$}}{\label{algo: AoG_column_start}
        $A \leftarrow$ (attributes of $G$ intersecting $A_{q_t}$)\;
        \For{$a \in A,\ c \in [g_2]$ where $B[a][c]$ is defined}{
            \textit{Cells}($c,a$) $\leftarrow$ set of cells in column/row $c$ of $G$, or the corresponding cells in a 1-D vs. 2-D comparison\;
            $x \leftarrow |S \,\cap\, \textit{Cells}(c,a)| \,-\, B[a][c]$\;
            \If{$x < 0$ \textbf{or} $x > 1$}{\label{algo: AoG_column_plus1}
                \Return \textbf{False}
            }
        }
        \For{$a \in A,\ c \in [g_2]$ where $B[a][c]$ is undefined}{
            \textit{Cells}($c,a$) $\leftarrow$ set of cells in column/row $c$ of $G$\;
            $B[a][c] \leftarrow |S \,\cap\, \textit{Cells}(c,a)|$\;
        }
        \Return \textbf{True}
    }\label{algo: AoG_column_end}

    \Fn{\findcover{$S, G$}}{\label{algo: AoG_find_hash_start}
        $R_{q_t}(G) \leftarrow$ set of cells in $G$ within $q_t$\;
        \For{$h \in \mathcal{H},\, c \in \mathcal{C}$}{\label{loop: hash}
            \If{$S[h,c] \,\subseteq\, R_{q_t}(G)$ 
                \textbf{and} 
                $|S[h,c]| \,\ge\, w_x$ (where $x$-D grid implies $w_x$)
                \textbf{and} 
                \columnconstraint{$S[h,c], G$}
            }{
                \Return $(h,c)$
            }
        }
        \textbf{Algorithm fails to find a hash pair.}
    }\label{algo: AoG_find_hash_end}

    $S_1[h,c] \leftarrow \{\,v \in \mathcal{G}_1 \mid h(v) = c\}$ for all $h \in \mathcal{H},\, c \in \mathcal{C}$\;
    $S_2[h,c] \leftarrow \{\,v \in \mathcal{G}_2 \mid h(v) = c\}$ for all $h \in \mathcal{H},\, c \in \mathcal{C}$\;
    $B[a_i][c_j] \leftarrow \textit{undefined}$ for all $a_i \in A_{q_t},\ c_j \in [g_2]$\;

    \For{each 2-D grid $G_{ij}$ with at least one dimension in $A_{q_t}$}{
        All fake users in $G_{ij}$ return \findcover{$S_2, G_{ij}$}\;
    }
    \For{each 1-D grid $G_i$ for dimension in $A_{q_t}$}{
        All fake users in $G_i$ return \findcover{$S_1, G_i$}\;
    }
\end{algorithm}

\begin{algorithm}
    \DontPrintSemicolon
    \caption{Heuristic Attack on Grid (HAoG)}
    \label{algo: AoG_heuristic}
    \textbf{Parameter}: $\mathcal{H}$ (the set of all hash functions). $\mathcal{C} = [g]$ (the set of hash keys). $A_{q_t}$ (attribute set where $q_t$ is concerned). $\mathcal{G}_1 = [g_1]$ (the set of cells in 1-D grids). $\mathcal{G}_2 = [g_2^2]$ (the set of cells in 2-D grids).\\
    $S_1[h,c] \leftarrow \{v | h(v) = c, v \in \mathcal{G}_1\} \forall h \in \mathcal{H}, c \in \mathcal{C}$. \\
    $S_2[h,c] \leftarrow \{v | h(v) = c, v \in \mathcal{G}_2\} \forall h \in \mathcal{H}, c \in \mathcal{C}$. \\
    \SetKwFunction{preference}{preference}
    \SetKwProg{Fn}{Function}{:}{}
    \Fn{\preference{$h,c,G$}}{
    \label{algo: AoG_heuristic_preference}
        $R_{q_t}(G) \leftarrow $ The set of cells in $G$ that are inside $q_t$. \\
        \If{$G$ is a 1-D grid}{
            \textbf{Return} $\left(\frac{|S_1[h,c] \cap R_{q_t}(G)| - |S_1[h,c]|}{g_1/g_2}, \frac{|S_1[h,c]|}{g_1/g_2}\right)$ \label{algo: AoG_heuristic_1D}\\
        }
        \textbf{Return} $\left(|S_2[h,c] \cap R_{q_t}(G)| - |S_2[h,c]|, |S_2[h,c]|\right)$.
    }
    
    \For{each grid $G$ having at least one dimension in $A_{q_t}$}{
        $(h,c) \leftarrow \arg\max_{(h,c)\in \mathcal{H} \times \mathcal{C}}$ \preference{$h,c,G$}. Tie breaks randomly.\\
        All fake users in $G$ return $(h,c)$
    }
\end{algorithm}

\subsection{Detection Algorithm}\label{sec: Detection Algorithm}
In Section \ref{sec: defense}, we introduce two detection algorithm based on hypothesis test. The detailed procedures for tree-based detection and grid-based detection is shown in Algorithms~\ref{algo: defense_tree} and \ref{algo: grid_based_defense}, respectively. 

\begin{algorithm}
    \DontPrintSemicolon
    \caption{Tree-based Detection}
    \label{algo: defense_tree}
    \textbf{Parameter}: $z_\alpha$ (Z-score of the significant threshold $\alpha$). $n$ (the number of bits in OUE). $N$ (number of users). $p,q$ (the probabilities used in OUE). $F(x)$ (the c.d.f. for distribution $Bin(n-1,q) + Bin(1,p)$).\\

    $f_I \leftarrow \frac{1 - \sqrt{\frac{1}{1 + z_\alpha^2}}}{2}$. \tcp*{This minimizes Equation \ref{eq: tree_defense_minimize}}
    $I^+,I^-$ be the number such that $F(I^+) = 1 - \frac{f_I}{2}$, $F(I^-) = \frac{f_I}{2}$. \\
    \If{more than $Nf_I + z_\alpha\sqrt{Nf_I(1-f_I)}$ users with the number of 1s more than $I^+$ or less than $I^-$}{
        \textbf{An Attack is Detected.}
    }
\end{algorithm}

\begin{algorithm}
    \DontPrintSemicolon
    \caption{Grid-based Detection}
    \label{algo: grid_based_defense}
    \textbf{Parameters}: $\alpha$ (significance threshold). $N$ (number of users). $|\mathcal{H}|$ (the number of hash functions). $L$ (maximum number of times a hash function is chosen among all hash functions).
    
    $F(x) \leftarrow$ c.d.f. of the maximum load of the $N$-balls-$|\mathcal{H}|$-bins problem, estimated by simulation.\\

    \If{$F(L) > 1 - \alpha$}{
        An attack is detected
    }
\end{algorithm}

\subsection{Adaptive Attack}\label{sec: Adaptive Attack}
Here, we introduce two adaptive attacks. The pseudocode for them is shown in Algorithms~\ref{algo: adaptive_tree} and \ref{algo: adaptive_grid}, respectively. 
\begin{algorithm}[hbt!]
    \DontPrintSemicolon
    \caption{Adaptive Attack on Tree (AAoT)}
    \label{algo: adaptive_tree}
    \textbf{Parameter}: $n$ the number of bits in OUE. $q = \frac{1}{e^\epsilon + 1}$.
    
    \For{each fake user $u_i$}{
        $X_i \leftarrow $ the data $u_i$ would have returned in Algo~\ref{algo: AoT}. \\
        $n_i \leftarrow $ the number of $1$s in $X_i$ \\
        $X \sim Bin(n-1,q) + Bin(1,\frac{1}{2})$ \\
        \If{$X > n_i$}{
            $X_i' \leftarrow $ randomly fills $X-n_i$ $0$s in $X_i$ into $1$s.\\
        }\Else{
            $X_i' \leftarrow $ randomly fills $n_i - X$ $1$s in $X_i$ into $0$s.\\
        }
        Fake user $u_i$ reports $X_i'$
    }
\end{algorithm}

\begin{algorithm}[]
    \DontPrintSemicolon
    \caption{Adaptive Attack on Grid (AAoG)}
    \label{algo: adaptive_grid}
    
    \textbf{Parameter}: $preference(\cdot)$ preference function in Algorithm \ref{algo: AoG_heuristic}. 
    $N,M$ the number of real, fake users. $\beta$ failing probability the attacker is willing to take. $\mathcal{H}$ the set of hash functions. $\mathcal{C} = [g]$ the set of hash keys. $t$ the threshold of grid defense. $A_{q_t}$ attribute set where $q_t$ is concerned. \\

    $L \leftarrow $ Maximum number $l$ s.t. $Pr[\text{the maximum load of randomly drawn $\frac{M}{l}$ bins from }(N+M)\text{-balls-}|\mathcal{H}|\text{-bins} < t - l] \geq 1 - \beta$ \label{algo: adaptive_grid_L} \\
    $G \leftarrow$ Set of 1-D and 2-D grids which have at least one attribute in $A_{q_t}$.\\
    $M_{G_i} \leftarrow$ Number of fake users in the grid $G_i$.\\
    Perform Stable Matching on $G$ and $\mathcal{H}$ with $G_i \in G$ having quota $M_{G_i}/L$ and $h \in \mathcal{H}$ having preference value $\max_{c \in \mathcal{C}}$ \preference{$h,c,G_i$} on $G_i \in G$.\\
    \For{$G_i \in G$}{
        $H \leftarrow$ Matched hash functions to $G_i$.\\
        \For{$h \in H$}{
            $c \leftarrow \max_{c \in \mathcal{C}}$ \preference{$h,c,G_i$}. \\ 
            Randomly select $L$ unused fake users in $G_i$. \\
            Those fake users submit $(h,c)$ to the server.
        }
    }
\end{algorithm}

\section{Proof of Theorem \ref{thm:tree_optimal}} \label{proof: tree_optimal}
\begin{proof}
Suppose \(A = \{a^{\mathrm{OPT}}_i\}_{i=1}^{|L|}\) is an optimal assignment that maximizes Eq.~\ref{eq:optimize_tree}, and let \(\hat{f}_A(v)\) be the frequency of \(v\) after applying Norm-Sub to \(A\). If there exist indices \(i, j\) with \(i < j\), \(a^{\mathrm{OPT}}_i < M_L\), and \(a^{\mathrm{OPT}}_j > 0\), we can construct a new assignment \(B = \{b_i\}_{i=1}^{|L|}\) that is better or at least as good as \(A\), by considering three cases:
\begin{enumerate}
\item \label{der: tree_prove_case1}
\textbf{If} \(\hat{f}_A(v_i) > 0\) and \(\hat{f}_A(v_j) > 0\):  
   Choose a small \(\varepsilon > 0\) and set  
   \[
       b_i = a^{\mathrm{OPT}}_i + \varepsilon,\quad
       b_j = a^{\mathrm{OPT}}_j - \varepsilon,\quad
       b_k = a^{\mathrm{OPT}}_k \;\;\forall\,k \neq i,j.
   \]
   Since \(\hat{f}_A(v_i)\) and \(\hat{f}_A(v_j)\) are both positive, the Norm-Sub threshold \(\delta\) (Eq.~\ref{eq:normsub}) remains unchanged when moving from \(A\) to \(B\). Therefore,
   \[
       \sum_{v \in L} c_v^{q_t, T} \hat{f}_B(v)
       -
       \sum_{v \in L} c_v^{q_t, T} \hat{f}_A(v)
       \;=\;
       (c_i^{q_t, T} - c_j^{q_t, T})
       \,\frac{\varepsilon}{N\bigl(\tfrac12 - q\bigr)}
       \;>\; 0,
   \]
   implying \(B\) is strictly better, contradicting \(A\)’s optimality.

\item \label{der: tree_prove_case2}
\textbf{If} \(\hat{f}_A(v_i) = 0\) and \(\hat{f}_A(v_j) > 0\):  
   Let \(b_i^\delta\) be the largest assignment to node \(i\) that keeps the Norm-Sub threshold \(\delta\) unchanged. If \(b_i^\delta < M_L\), choose a small \(\varepsilon > 0\) and set  
   \[
       b_i = b_i^\delta + \varepsilon,\quad
       b_j = a^{\mathrm{OPT}}_j - \varepsilon,\quad
       b_k = a^{\mathrm{OPT}}_k \;\;\forall\,k \neq i,j.
   \]
   By the same argument as in case~\ref{der: tree_prove_case1}, this yields a better assignment than \(A\). Otherwise, let  
   \[
       b_i = M_L,\quad 
       b_k = a^{\mathrm{OPT}}_k
       \;\;\forall\,k \neq i,
   \]
   producing an assignment \(B\) at least as good as \(A\).

\item \label{der: tree_prove_case3}
\textbf{If} \(\hat{f}_A(v_j) = 0\):  
   Set \(b_j = 0\) and \(b_k = a^{\mathrm{OPT}}_k\) for all \(k \neq j\), yielding \(B\) that is at least as good as \(A\).
\end{enumerate}
By iteratively applying cases~\ref{der: tree_prove_case2} and \ref{der: tree_prove_case3}, any optimal assignment \(A\) can be converted into an equally good assignment \(B\) where exactly one index \(i\) satisfies \(0 \le b_i < M_L\), all indices \(j < i\) satisfy \(b_j = M_L\), and all indices \(k > i\) satisfy \(b_k = 0\).
\end{proof}

\section{$\Theta\left(|L|^2\log{|L|}\right)$ Search Algorithm for Optimal Assignment} \label{sec:efficientsearch}
Recall from Section~\ref{sec:aot} that the algorithm aims to find an optimal assignment maximizing Eq.~\ref{eq:optimize_tree}. According to Theorem~\ref{thm:tree_optimal}, there exists an optimal assignment of a specific form. Let \(L\) be the set of nodes in a tree layer, and denote \(\mathbf{L} = \lvert L\rvert\). We present an algorithm that locates an optimal assignment for \(L\) in \(\Theta(\mathbf{L}^2 \log \mathbf{L})\) time.

Because Norm-Sub on a length-\(\mathbf{L}\) frequency vector can be performed in \(\Theta(\mathbf{L}\log\mathbf{L})\) time (by sorting and accumulating frequencies from top to bottom), our key objective is to evaluate potentially optimal assignments while consistently maintaining Norm-Sub’s threshold.

\paragraph{Base Assignments.}
Define the \(i\)-th base assignment as \(A_i = \{a_{ij}\}_{j=0}^{\mathbf{L}}\), where
\[
a_{ij} 
=\; 
\begin{cases}
    M_L, & \text{if } i < j,\\
    0, & \text{otherwise.}
\end{cases}
\]
We introduce a “trailing” parameter \(t\) for each base assignment \(0 \le t < M_L\). The algorithm iterates over all base assignments \(A_i\). For each \(A_i\):
1. It first computes Norm-Sub to obtain the threshold \(\delta\) and the value \(v\) of \(A_i\).
2. It then considers adjusting \(t\) (the trailing) to explore potential improvements in \(\delta\).

\paragraph{Maintaining Norm-Sub.}
Let \(F_i\) be the frequency vector when applying the \(i\)-th base assignment. Sort \(F_i\) in ascending order. The Norm-Sub threshold \(\delta\) always appears immediately before some frequency \(f_k \in F_i\). To raise \(\delta\) to \(f_k\), we increase \(t\) to \(t'\) so that each relevant \(f_i\) becomes \(f_i'\). The new value \(v'\) changes by
\[
c_i\,\bigl(f_i' - f_i\bigr)
\;-\;
\sum_{h \in [\mathbf{L}]} 
    c_h \,\mathbbm{1}_{f_h \,\geq\, \delta}\,
    \bigl(\delta - f_k\bigr).
\]
The second term can be updated efficiently using suffix sums. By iteratively raising \(\delta\) to the next frequency until \(t = M_L\), we find the best value \(v\) for some trailing \(t\) in the \(i\)-th base assignment in \(\Theta(\mathbf{L}\log \mathbf{L})\) time.

\paragraph{Time Complexity.}
Because there are \(\mathbf{L}\) possible base assignments \(A_i\), and each takes \(\Theta(\mathbf{L}\log \mathbf{L})\) to evaluate via this procedure, the total time is \(\Theta(\mathbf{L}^2 \log \mathbf{L})\).

\begin{algorithm}[hbt!]
    \DontPrintSemicolon
    \caption{AoT: Find Best Assignment Efficiently}
    \textbf{Parameter}: Coefficient in Eq. \ref{eq:optimize_tree} in descending order $C$. Frequency from real data (assuming uniformly distributed) $F$. The number of real user $N$. The number of fake user $M$. The probability in OUE $q$.

    $\mathbf{L} \leftarrow  |C|$ \\
    unit $\leftarrow \frac{1}{(N+M)(0.5-q)}$\\
    $A^{OPT}, v^{OPT} \leftarrow \emptyset, -1$. \\

    \For{$i\in [\mathbf{L}]$}{
        $A_i \leftarrow$ $i$-th base assignment.\\
        $F_i \leftarrow $ Frequency estimated after OUE using $F$ and $A_i$.\\
        $\delta, v \leftarrow$ The threshold in Norm-Sub and the value in Eq. \ref{eq:optimize_tree}.\\
        
        $A_i^{OPT}, v_i^{OPT} \leftarrow A_i, v$.\\
        $F_{i,\text{sorted}} \leftarrow $ Sorting $F_i$ in ascending order.\\
        $F_{\text{ss}}, C_{\text{ss}} \leftarrow $ Suffix-sum of $F_{i,\text{sorted}}$, $C$ using $F_{i,\text{sorted}}$.\\
        $f_i \leftarrow F_i[i]$. \\
        $t \leftarrow 0$. \\
        \If{$f_i < \delta$}{
            $t \leftarrow (\delta - f_i)/\text{unit}$.\\
            $f_i \leftarrow \delta$.\\
        }
        \For{$k \in [\mathbf{L}]$}{
            \If{$F_{i,\text{sorted}}[k] \leq \delta$}{
                \textbf{Continue.}
            }
            req\_unit $\leftarrow (1 - F_{\text{ss}}[k] + F_{i,\text{sorted}}[k](\mathbf{L} - k + 1) - f_i)/\text{unit}$.\\
            max\_unit $\leftarrow \min(\text{req\_unit}, M - t)$.\\
            $\Delta \delta \leftarrow [F_{\text{ss}}[k] - \delta(\mathbf{L} - k + 1) + \text{max\_unit}\cdot \text{unit} + f_i - 1]/(\mathbf{L} - i + 1)$.\\
            $v \leftarrow v + \text{max\_unit}\cdot \text{unit}\cdot C[k] - (C_{\text{ss}}[k] + C[k])\Delta\delta$.\\
            $t \leftarrow t + \text{max\_unit}$.\\
            $f_i \leftarrow \text{max\_unit}\cdot \text{unit}$.\\
            $\delta \leftarrow \delta + \Delta\delta$.\\
            \If{$v > v_i^{OPT}$}{
                $A_i^{OPT}[k], v_i^{OPT} \leftarrow t, v$.
            }
            \If{req\_unit $>$ max\_unit}{
                \textbf{Break}.
            }
        }
        \If{$v_i^{OPT} > v^{OPT}$}{
            $A^{OPT}, v^{OPT} \leftarrow A_i^{OPT}, v_i^{OPT}$.
        }
    }

    \textbf{Return} $A^{OPT}$
\end{algorithm}

\section{Proof of Theorem \ref{thm:fullycon}}\label{proof:fullycon}
\begin{proof}
    In the response matrix generation ~\cite{yang2020answeringmultidimensionalrangequeries}, there are no frequency for any cell $s$ outside of $q_t$. As a result, the response matrices also have $0$ for entries outside of $q_t$ (refer to line 9 in Algorithm 1 in ~\cite{yang2020answeringmultidimensionalrangequeries}). Similarly, since multidimensional queries are estimated using the same algorithm with the response matrices, the final estimation will also be $0$ outside $q_t$. Therefore, as the sum of the matrices is $1$ and entries outside of $q_t$ is all $0$, $\mathcal{M}_{X_{real} \cup X_{fake}}(q_t) = 1$.
\end{proof}

\section{Proof of Theorem \ref{thm: grid_constraint}} \label{proof:grid_optimal}
\begin{proof}
    Observe that in any 2-D grid $G$, if there exist $w$ cells in $q_t$ whose frequencies exceed those of every cell outside $q_t$ by more than $1/w$ immediately before Norm-Sub, the frequencies in $q_t$ will be fully concentrated after Norm-Sub.

Assuming uniformly distributed real users, a cell without fake user responses (i.e., not chosen by fake users) in OLH has frequency
\[
f_0 = \frac{(1-\rho)}{g_2^2} \;-\; \frac{\rho}{\,g\!\bigl(\tfrac{1}{2} - \tfrac{1}{g}\bigr)},
\]
while a cell chosen by all fake users has frequency
\[
f_1 = f_0 \;+\; \frac{\rho}{\,\tfrac{1}{2} - \tfrac{1}{g}\,}.
\]

Next, we consider the consistency step. The column constraint ensures that the sum of frequencies in the same column across different tables cannot exceed $f_1 - f_0$, since the maximum difference in chosen cells is 1. Consequently, the minimum frequency of a chosen cell after consistency is:
\[
f_1' \;\ge\; f_1 \;+\; 2\,\frac{\bigl(f_0 - f_1\bigr)}{g_2}
\Bigl(1 - \frac{\tfrac{1}{g_2}}{\tfrac{d-1}{g_2} + \tfrac{g_2}{g_1}}\Bigr),
\]
where the factor 2 arises because consistency applies once to each column and once to each row.

In contrast, a cell $c$ outside $q_t$ lies in a column or row that does not intersect $q_t$. By the inclusive constraint, there are no chosen cells in every corresponding column (or row), so one consistency adjustment is 0, and the other is limited by the column constraint to at most $f_1 - f_0$. Hence, the maximum frequency of a cell not chosen after consistency is:
\[
f_0' \;\le\; f_0 \;+\; \frac{f_1 - f_0}{g_2}
\Bigl(1 - \frac{\tfrac{1}{g_2}}{\tfrac{d-1}{g_2} + \tfrac{g_2}{g_1}}\Bigr).
\]

Because
\[
f_1' - f_0' \;\ge\; 
\bigl(f_1 - f_0\bigr)\!\Bigl(
1 - \frac{3}{g_2}\Bigl(1 - \frac{\tfrac{1}{g_2}}{\tfrac{d-1}{g_2} + \tfrac{g_2}{g_1}}\Bigr)
\Bigr)
\;=\; \frac{1}{w_2},
\]
and there are more than $w_2$ chosen cells in every 2-D grid by the size constraint, each 2-D grid is fully concentrated in $q_t$ after Norm-Sub.

A similar argument holds for 1-D grids. During consistency, both chosen and non-chosen cells are adjusted once. Thus, in 1-D:
\[
f_1' - f_0' \;\ge\; \bigl(f_1 - f_0\bigr)\!\Bigl(
1 - \frac{2\,g_2}{g_1}\Bigl(1 - \frac{\tfrac{g_2}{g_1}}{\tfrac{d-1}{g_2} + \tfrac{g_2}{g_1}}\Bigr)
\Bigr)
\;=\; \frac{1}{w_1}.
\]
Consequently, the 1-D grid also becomes fully concentrated in $q_t$ after Norm-Sub.
\end{proof}

\section{PRISM Breaks Parallel Composition Theorem}\label{appendix:prism}
In PRISM~\cite{prism}, the authors propose a randomized mechanism called Range-based Randomized Response (RRR), which they claim guarantees $\epsilon$-local differential privacy ($\epsilon$-LDP). We show via a simple counterexample that this mechanism does not, in fact, provide $\epsilon$-LDP.

RRR works as follows. Consider a 1-dimensional domain $\mathcal{D} = [0, \dots, d - 1]$. For each user, the mechanism produces a $|\mathcal{D}|$-bit vector, where the $i$-th bit denotes whether the user’s data is in $[0, i]$. This bit vector is then perturbed using randomized response with parameters $p = \frac{e^\epsilon}{e^\epsilon + 1}$ and $q = \frac{1}{e^\epsilon + 1}$. According to the authors, this procedure satisfies $\epsilon$-LDP.

However, because different bits in the vector share overlapping information, the parallel composition theorem cannot be applied directly. Instead, the mechanism delivers $(d\epsilon)$-LDP. As a simple illustration, consider the domain $\mathcal{D} = [0, 1, 2]$ with $d = 3$.

\textbf{Counterexample:}  
Suppose the user’s data is either $0$ or $2$, and the perturbed outcome is $(0, 0, 0)$:
\begin{itemize}
    \item With true data $0$, the unperturbed vector is $(1, 0, 0)$, which becomes $(0, 0, 0)$ with probability $p^2 q$.
    \item With true data $2$, the unperturbed vector is $(1, 1, 1)$, which becomes $(0, 0, 0)$ with probability $q^3$.
\end{itemize}
The ratio of these probabilities is
\[
\frac{\Pr[(0, 0, 0)\,\mid\,\text{data} = 0]}{\Pr[(0, 0, 0)\,\mid\,\text{data} = 2]} 
= \frac{p^2 q}{q^3} 
= \frac{p^2}{q^2} 
= e^{2\epsilon},
\]
which exceeds $e^\epsilon$ and thus violates $\epsilon$-LDP. Moreover, the original paper’s link to a proof of $\epsilon$-LDP is no longer functional.

Although the mechanism may satisfy LDP according to the basic composition theorem, its low estimation variance critically depends on assuming the parallel composition theorem applies—which we have shown is invalid. From the basic composition theorem, the privacy budget of PRISM scales with the domain size, which can be large for most real-world datasets. \textbf{In other words, their experimental findings rely on a severely underestimated privacy budget, effectively undermining their conclusions.} Consequently, we do not report our attack results on PRISM, as it lacks a solid theoretical basis.

\section{More Experiment for AHEAD}\label{sec:even_more_tree}
\subsection{Different $\epsilon$ and $\rho$}
Figure A.1$\sim$A.14 in the Appendix show the experimental results on all dataset with various parameter settings. The result demonstrates that our attacks perform consistently well, achieving high responses in almost every parameter setting.
\subsection{Existing Defense}
The result in Figure A.15$\sim$A.28 and G.60$\sim$G.67 in the Appendix demonstrates that LDPRecover is not a suitable defense for LDP range query estimation. LDPRecover incurs large variance in the query estimation as they obstruct the decomposition of tree nodes. Especially in Salaries, Bfive and Sythetic dataset where real response is high, LDPRecover produces enormous variance, making AHEAD with LDPRecover unusable even when no attack is launched. On the other hand, since Loan dataset contains a more flat distribution, its response does not get strongly affected by the obstruction of decomposition of tree nodes.
\subsection{Detection}\label{exp:tree_detection_more}
In Figure A.29 in the Appendix, MGA is detected with close to 100\% detection rate in almost every setting and dataset, while AoT is detected occasionally, with 40 to 50\% of detection rate. On the other hand, AAoT consistently dodges the detection with a detection rate close to detection rate on real data (LDP).

\section{More Experiment on HDG}\label{sec:even_more_grid}
\subsection{Different $\epsilon$, $\rho$ and the number of times of post-processing}
In Figure G.1$\sim$G.56 in the Appendix, we present the results of attacks on the grid-based approach across various datasets and parameter settings. The results demonstrate that AoG and HAoG outperform MGA in most cases. However, AoG and HAoG in extremely rare case may exhibit zero response after the first round of post-processing. This occurs because when AoG fails to identify suitable hash pairs, it defaults to HAoG, which selects hash pairs without considering consistency. Specifically, HAoG might unforntunately choose hash pairs for 1-D and 2-D grids where the domains of the cells within the support of the hash pairs do not overlap. Consequently, the response matrix estimates the response for these domains as zero since the frequencies of non-overlapping cells become zero during response matrix generation. Nonetheless, such cases are infrequent, and the attack regains effectiveness after subsequent post-processing.

\subsection{Detection}
In Figure G.57 in the Appendix, the detection rate demonstrates that the detection algorithm can successfully detect attacker using maximum load. However, AAoG with $\beta = 10\%$ can dodge the detection around 90\% of the time. The result displays that attacker using AAoT can control the risk of getting caught accurately using simulation to calculate $l$.

\section{Defense Result via LDPRecover}\label{sec: Detection Result via LDPRecover}

\begin{figure}[bt!]
    \centering
      \begin{minipage}{0.24\textwidth}
    \includegraphics[width=\linewidth]{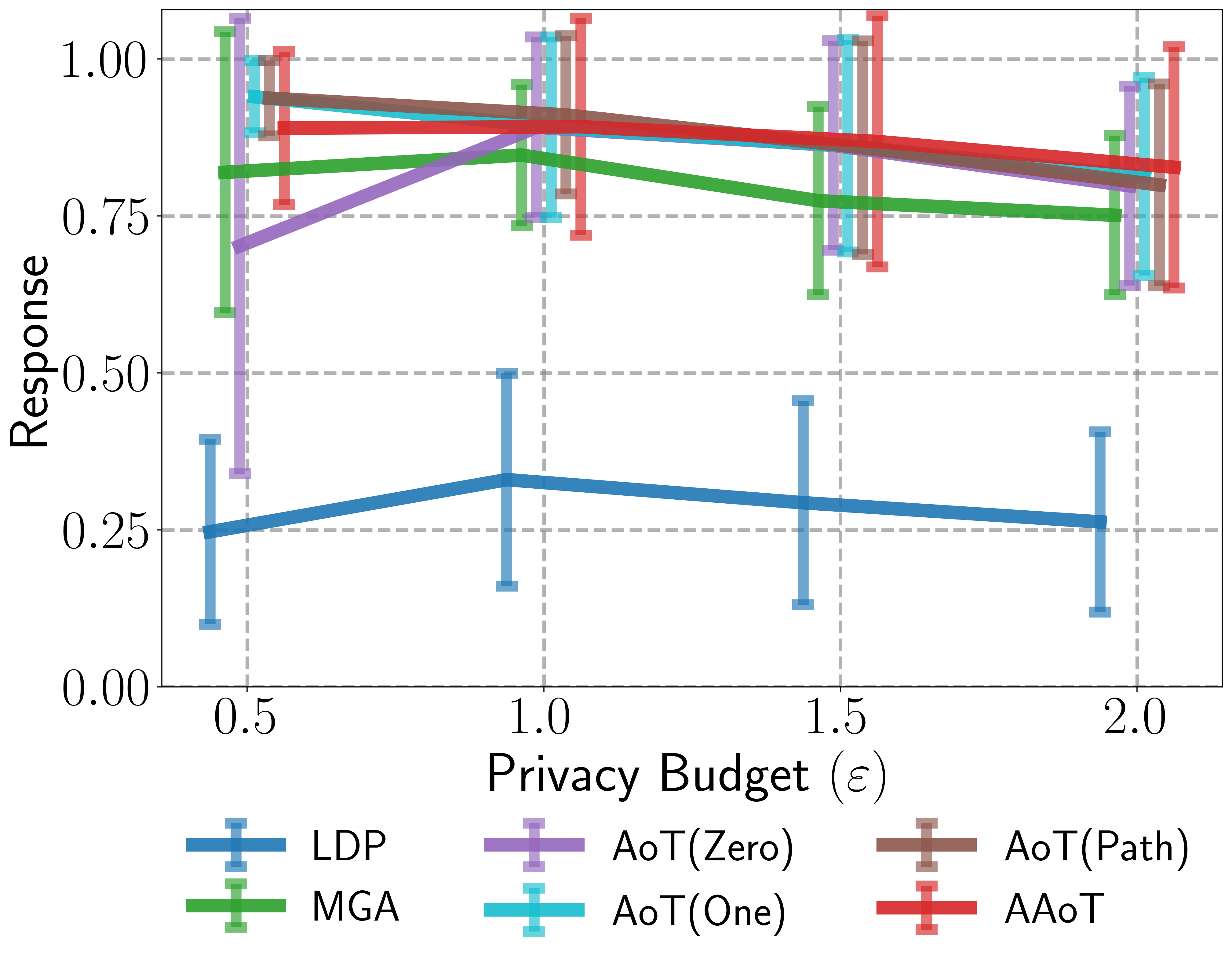}
    \subcaption{LDPRecovery on AHEAD}
    \label{fig:tree_ldpr}
  \end{minipage}\begin{minipage}{0.24\textwidth}
    \includegraphics[width=\linewidth]{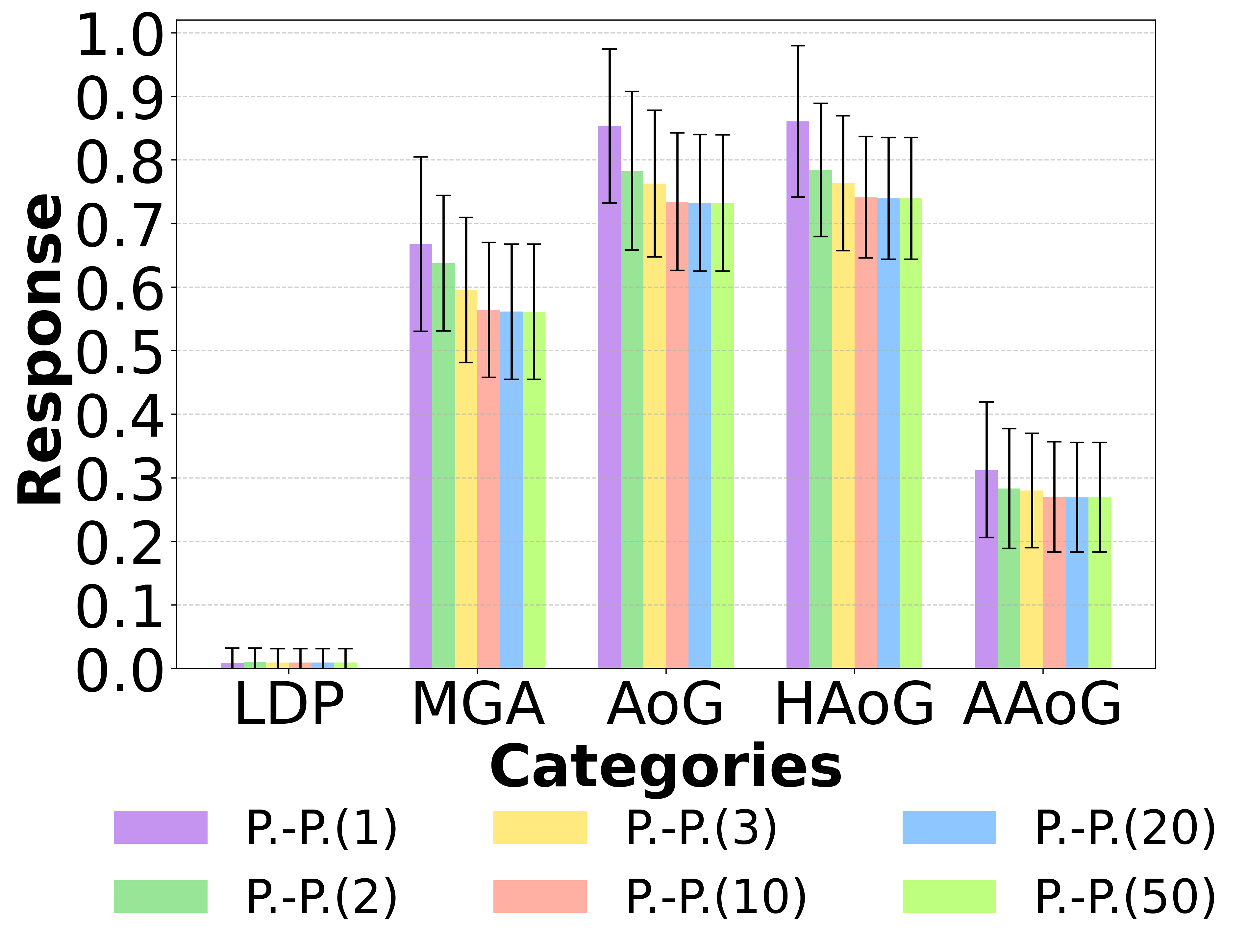}
    \subcaption{LDPRecovery on HDG}
    \label{fig:grid_ldpr}
    \end{minipage}
    \caption{The response with LDPRecover on BlackFriday dataset of different methods with $\epsilon = 1.0$, $\rho = 0.1$.}
\end{figure}

\textbf{Tree.} 
we observe that LDPRecover is not an effective defense for the AHEAD protocol (see Figure~\ref{fig:tree_ldpr}). Its influence is minimal when the real response is small, and negligible when the real response from target query is not close to 0, substantially limiting its usefulness. Additionally, LDPRecover introduces substantial estimation variance into LDP, leading to a high performance cost. This large variance mainly arises from its disruption of node decomposition. By normalizing frequencies among nodes, LDPRecover often suppresses node decomposition by lowering their frequencies, thereby preventing the tree from estimating frequencies at finer granularity.

\textbf{Grid.} we observe that LDPRecover imposes minimal impact on any of our proposed methods or MGA (see Figure~\ref{fig:grid_ldpr}). Its effect is analogous to an additional round of post-processing. Because LDPRecover lacks insight into the attacker’s strategy, it simply treats all positive frequencies as targeted items and applies normalization. In practice, this normalization step behaves like a consistency procedure, lowering frequencies inside the target range while raising frequencies outside it. However, our attack methods—especially AoG—are carefully tailored to remain robust against the first consistency step, rendering LDPRecover’s defense largely ineffective.

\end{document}